\documentclass{aa}

\usepackage{graphicx}
\usepackage{bm}
\usepackage{txfonts}
\usepackage{mathrsfs}
\usepackage{natbib}

\newcommand{\met}{[\element{Fe}/\element{H}]}

\newcommand{\caii}{\ion{Ca}{II}}
\newcommand{\nravespectra}{574,630 }
\newcommand{\nravestars}{483,330 }
\newcommand{\ndrfourstars}{457,588 }
\newcommand{\nmpcandidate}{3174 }
\newcommand{\nvmpcandidate}{877 }
\newcommand{\ncalibration}{153 }
\newcommand{\nempcandidate}{43 }

\usepackage{color}
\usepackage{amstext}

\begin{document}

\title{Very metal-poor stars observed by the RAVE survey}

\author{
G.~Matijevi\v c \inst{1} \and 
C.~Chiappini \inst{1} \and 
E.~K.~Grebel \inst{2}  \and 
R.~F.~G.~Wyse \inst{3} \and 
T.~Zwitter \inst{4} \and
O.~Bienaym\' e \inst{5} \and 
J.~Bland-Hawthorn \inst{6} \and 
K.~C.~Freeman \inst{7} \and 
B.~K.~Gibson \inst{8} \and 
G.~Gilmore \inst{9} \and 
A.~Helmi \inst{10} \and 
G.~Kordopatis \inst{1} \and 
A.~Kunder \inst{1, 11} \and
U. Munari \inst{12} \and
J.~F.~Navarro \inst{13}\thanks{CIfAR Senior Fellow} \and 
Q.~A.~Parker \inst{14} \and 
W.~Reid \inst{15, 16} \and 
G.~Seabroke \inst{17} \and 
A.~Siviero \inst{18} \and
M.~Steinmetz \inst{1} \and 
F.~Watson \inst{19}}

\institute{
Leibniz-Institut f\"ur Astrophysik Potsdam (AIP),
              An der Sternwarte 16, 14482 Potsdam, Germany
              \email{gmatijevic@aip.de}              
\and Astronomisches Rechen-Institut, Zentrum f\"ur Astronomie der Universit\"at Heidelberg, M\"onchhofstr. 12-14, 69120 Heidelberg, Germany                        
\and Department of Physics and Astronomy, Johns Hopkins University, 3400 N. Charles St, Baltimore, MD 21218, USA
\and Faculty of Mathematics and Physics, University of Ljubljana, 1000 Ljubljana, Slovenia
\and Observatoire astronomique de Strasbourg, Universite de Strasbourg, CNRS, UMR 7550, 11 rue de l'Universite, F-67000 Strasbourg, France
\and Sydney Institute for Astronomy, School of Physics A28, University of Sydney, NSW 2006, Australia
\and Research School of Astronomy and Astrophysics, Australian National University, Cotter Rd, Weston, ACT 2611, Australia
\and E.~A.~Milne Centre for Astrophysics, University of Hull, Hull, HU6 7RX, United Kingdom 
\and Institute of Astronomy, University of Cambridge, Madingley Road, Cambridge CB3 0HA, UK
\and Kapteyn Astronomical Institute, University of Groningen, P.O. Box 800, NL-9700 AV Groningen, The Netherlands
\and Saint Martin's University, Old Main, 5000 Abbey Way SE, Lacey, WA 98503, USA
\and INAF Astronomical Observatory of Padova, 36012 Asiago (VI), Italy
\and Department of Physics and Astronomy, University of Victoria, Victoria, BC, Canada V8P 5C2
\and Department of Physics, Chong Yuet Ming Physics Building, The University of Hong Kong, Hong Kong
\and Department of Physics and Astronomy, Macquarie University, Sydney, NSW 2109, Australia
\and Western Sydney University, Locked Bag 1797, Penrith South DC, NSW 1797, Australia
\and Mullard Space Science Laboratory, University College London, Holmbury St Mary, Dorking, RH5 6NT, UK
\and Dipartimento di Fisica e Astronomia Galileo Galilei, Universita’ di Padova, Vicolo dell’Osservatorio 3, I-35122 Padova, Italy
\and Australian Astronomical Observatory, North Ryde, NSW 2113, Australia
}      

\abstract{ Metal-poor stars trace the earliest phases in the chemical enrichment of the Universe. They give clues about the early assembly of the Galaxy as well as on the nature of the first stellar generations. Multi-object spectroscopic surveys play a key role in finding these fossil records in large volumes. Here we present a novel analysis of the metal-poor star sample in the complete Radial Velocity Experiment (RAVE) Data Release 5 catalog with the goal of identifying and characterizing all very metal-poor stars observed by the survey. Using a three-stage method, we first identified the candidate stars using only their spectra as input information. We employed an algorithm called t-SNE to construct a low-dimensional projection of the spectrum space and isolate the region containing metal-poor stars. Following this step, we measured the equivalent widths of the near-infrared $\caii$ triplet lines with a method based on flexible Gaussian processes to model the correlated noise present in the spectra. In the last step, we constructed a calibration relation that converts the measured equivalent widths and the color information coming from the 2MASS and WISE surveys into metallicity and temperature estimates. We identified $\nvmpcandidate$ stars with at least
a $50\%$ probability of being very metal-poor $(\met < -2\,\mathrm{dex})$, out of which $\nempcandidate$ are likely extremely metal-poor $(\met < -3\,\mathrm{dex})$. The comparison of the derived values to a small subsample of stars with literature metallicity values shows that our method works reliably and correctly estimates the uncertainties, which typically have values $\sigma_{\mathrm{[Fe/H]}} \approx 0.2\,\mathrm{dex}$. In addition, when compared to the metallicity results derived using the RAVE DR5 pipeline, it is evident that we achieve better accuracy than the pipeline and therefore more reliably evaluate the very metal-poor subsample. Based on the repeated observations of the same stars, our method gives very consistent results. We intend to study the identified sample further by acquiring high-resolution spectroscopic follow-up observations. The method used in this work can also easily be extended to other large-scale data sets, including to the data from the Gaia mission and the upcoming 4MOST survey.}


\maketitle

\section{Introduction}
Metal-poor stars have been extensively used in stellar and Galactic archaeology studies over the past few decades \citep[e.g.,][]{2005ARA&A..43..531B}. They offer an insight into the nucleosynthesis in the early Galaxy \citep{2013RvMP...85..809K} and subsequent metal enrichment by  supernovae of type II \citep{2004ARA&A..42...79B, 2010AN....331..474F, 2013AN....334..595C}, with special attention drawn toward carbon-enhanced metal-poor stars \citep{2012ApJ...744..195C, 2013A&A...552A.107S, 2016A&A...588A..37H} and other evolved stars. The most metal-poor stars in the halo (with metallicities $\met<-4.0$) are also believed to be the oldest\footnote{In the bulge, very old stars are found at higher metallicities of  $\met\approx-1.5$ \citep{2011Natur.472..454C}.}. Therefore, they allow us to study the conditions in the young Universe in which they were born \citep{2013pss5.book...55F}. In addition, the Galaxy's accretion history can be better understood by observing metal-poor stars in the Galactic halo and surrounding dwarf galaxies \citep[and references therein]{2009ARA&A..47..371T, 2010A&A...513A..34S}.

Given the broad range of studies that can be conducted with metal-poor stars, the systematic search for them began early on \citep[][and the references therein]{2005ARA&A..43..531B}. Great success was achieved by the HK survey \citep{1985AJ.....90.2089B, 1992AJ....103.1987B} and the Hamburg/ESO survey \citep[HES,][]{1996A&AS..115..227W, 2006ApJ...652.1585F, 2008A&A...484..721C}, yielding more than 1000 very metal-poor stars ($\met < -2.0$) and a few hundred extremely metal-poor stars ($\met<-3.0$). Higher resolution follow-up spectroscopy revealed that several stars discovered by the HES are among the most metal-poor stars known \citep{2002Natur.419..904C, 2005Natur.434..871F}. Following their success, several wide-field sky surveys \citep[][among others]{2007PASA...24....1K, 2010ApJ...724L.104F, 2013AJ....145...13A, 2014MNRAS.445.4241H, 2015A&A...579A..98A, 2015PASJ...67...84L} enabled discoveries of new extremely metal-poor stars, including the current metal-poor record holder SM 0313-6708 \citep{2014Natur.506..463K} with an upper limit estimate of $\met=-7.3$. Currently, there are several hundred known stars with metallicities of $\met < -2.5$ \citep{2013ApJ...762...25N} and just over 20 stars with $\met<-4.0$ \citep{2015ApJ...810L..27F}. More candidates have since been identified \citep{2015ApJ...809..136P}. The metallicity distribution function (MDF) of the metal-poor end is relatively unconstrained because only a few metal-poor stars are known, and any additional discoveries of new candidates advance our knowledge of the detailed shape of the complete metal-poor MDF.

An earlier study of very metal-poor stars that were found with the Radial Velocity Experiment (RAVE) survey was performed by \citet{2010ApJ...724L.104F} on a sample of $\sim200,000$ stars available in the survey at the time. The authors identified metal-poor candidates based on the atmospheric parameters derived from the RAVE spectra. They obtained high-resolution observations for a subset of 112 stars and used them for calibrating the remainder of the selected candidate sample. With the RAVE observations meanwhile completed, this study aims to find \textit{all} very metal-poor star in the final RAVE sample and therefore to extend the previous work by introducing a more robust three-stage analysis of the spectra and the calibration relation.

This article is structured as follows. Section \ref{sect_obs} gives an overview of the RAVE spectra acquisition and reduction procedures and highlights the importance of having a separate processing pipeline for very metal-poor stars in addition to the main pipeline, which is successfully used on the large majority of other RAVE stars. In Section \ref{sect_id} we present a method for isolating the very metal-poor sample, followed by a review of the method used for line profile modeling in Section \ref{sect_gpmodel}. Its output along with stellar color information is used to derive metallicity values of candidate stars (Section \ref{sect_met}). In Section \ref{sect_prop} we analyze the calibrated sample and compare our measurements to those RAVE stars that have also been observed by other studies targeting metal-poor stars. In this way, we obtain a list of all metal-poor candidates in RAVE to be confirmed by high-resolution follow-up observations. We conclude
with a discussion of the potential impact our method might have on future massive spectroscopic surveys.

\section{Observations and spectral analysis}
\label{sect_obs}

The RAVE survey \citep{2006AJ....132.1645S} started collecting first spectra in 2003 and ran until 2013. Its main goal was to  measure radial velocities and atmospheric parameters of predominantly disk stars \citep{2008AJ....136..421Z, 2011AJ....141..187S}. During this period, it gathered \nravespectra spectra of \nravestars stars using the Six Degree Field (6dF) multi-object spectrograph installed on the 1.2 m UK Schmidt Telescope of the Australian Astronomical Observatory. The input catalog for the survey was initially based on the Tycho-2 \citep{2000A&A...355L..27H} and SuperCOSMOS \citep{2001MNRAS.326.1279H} catalogs with derived and direct $I$-band photometry, respectively, but was later switched to 2MASS positions \citep{2003tmc..book.....C} and the $I$-band photometry from the DENIS catalog \citep{1997Msngr..87...27E}. Stars selected for observation were chosen to be in the magnitude range between $9 < I_\mathrm{DENIS} < 13$ and to avoid the Galactic plane and the bulge. The spectral range includes the near-infrared $\caii$ triplet region and covers the wavelengths of $8410\--8795\, \mathrm{\AA}$.

After the extraction of the spectra from the CCD image, the sky emission lines are subtracted, the continua of spectra are normalized, and the spectra are shifted to the rest-frame velocity according to individual radial velocity measurements. The typical resolving power of the spectra is around $R\sim 7500$ and the median signal-to-noise ratio per pixel value, $S/N$, is $\sim 50$.  The latest version of the stellar parameters catalog provided by \citet{2017AJ....153...75K} as the fifth data release (DR5) includes effective temperatures, surface gravities, and metallicities of \ndrfourstars stars. The processing pipeline used to derive the parameters employs the  MATISSE \citep[a projection algorithm,][]{2006MNRAS.370..141R}, DEGAS (a classification algorithm), and a library of precomputed synthetic spectra to extract the parameters from the observed spectra. It excludes the central regions of the $\caii$ lines because difficulties arise from modeling those lines \citep{2011A&A...535A.106K}. In addition to the information provided by the parameter estimation pipeline, \citet{2011AJ....142..193B} computed individual abundances for \ion{Al}, \ion{Fe}, \ion{Mg}, \ion{Ni}, \ion{Si}, and \ion{Ti}\ for a subset of cooler stars with higher quality spectra. The mean uncertainties for these measurements are in the range of $\sim 0.2\,\mathrm{dex}$.  Distances to the observed stars as the missing piece to obtain the full six-dimensional kinematical information were first computed by \citet{2010A&A...511A..90B} and later revised by \citet{2010A&A...522A..54Z}, \citet{2011A&A...532A.113B}, and \citet{2014MNRAS.437..351B}. Most of the stars in the sample were also observed by the 2MASS survey, so there are $J$, $H$, and $K_S$ band apparent magnitudes available as well as all four bands in the WISE catalog \citep{2010AJ....140.1868W}.

\begin{figure}
\centering
\resizebox{\hsize}{!}{\includegraphics{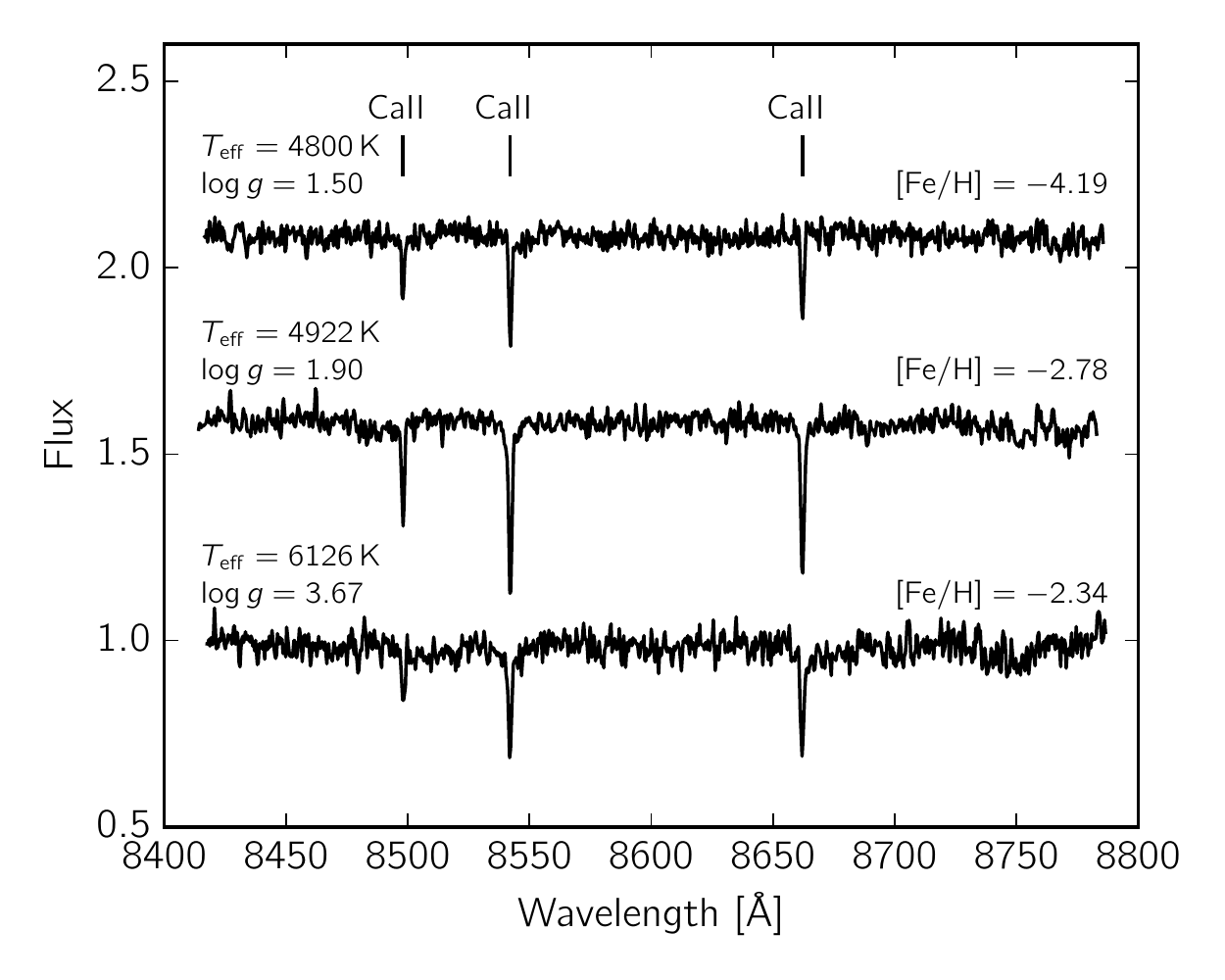}}
\caption{Three examples of very metal-poor RAVE spectra with typical $S/N$ values of $\sim 50$. The top spectrum belongs to the star with the lowest metal abundance in the catalog, CD -38 245. The parameter values for the top spectrum were adopted from \citet{2013A&A...551A..57H}, the parameters for the middle spectrum from \citet{2013MNRAS.434.1681C}, and for the bottom spectrum we used parameters from \citet{2013MNRAS.429..126R}. Note the degeneracy between the shape of the $\caii$ lines of cooler giants with lower metallicities and warmer turn-off stars with higher metallicities.}
\label{fig_mp_spectra}
\end{figure}

The majority of the spectra are dominated by the $\caii$ lines located at $8498\, \mathrm{\AA}$, $8542\, \mathrm{\AA}$, and $8662\, \mathrm{\AA}$. In addition to these lines, numerous other metallic lines are present in the spectra. Hotter stars (A and earlier types) exhibit strong hydrogen lines, while cooler M-type giants have a strong TiO molecular band. These features enable the parameter estimation pipeline to determine the values of the atmospheric parameters across the broad range in temperature, gravity, and metallicity. However, with increasingly lower metallicity ($\met < -2.0$), metallic lines in addition to the $\caii$ triplet become indistinguishable from the background noise at typical $S/N$ values. At lowest metallicities ($\met<-3.0$), other metallic lines are undetectable even at higher $S/N$ values, and only the three weak $\caii$ lines are clearly distinguishable (see examples in Fig. \ref{fig_mp_spectra}). As a consequence, the RAVE stellar parameter pipeline has difficulties delivering accurate parameters in this part of the parameter space. In many cases, the pipeline does not converge or it finds more than one solution, after which the average is taken as the final guess. Another reason for the poor performance in the very metal-poor regime is the fact that local thermodynamic equilibrium (LTE) spectral models cannot describe the $\caii$ spectral line sufficiently well. \citet{2010A&A...513A..34S} studied the discrepancies between the observations and the models and derived a correction term. However, this correction cannot be simply applied to the values derived by the parameter estimation pipeline. The correction itself is strongly dependent on the effective temperature and metallicity as an input, which are challenging to obtain spectroscopically without large uncertainties in the metal-poor regime. A more detailed explanation is given in Section \ref{sect_met}.

\section{Identification of metal-poor FGK stars}
\label{sect_id}

Among the main goals of the RAVE survey are the measurements of the stellar radial velocities and the determination of atmospheric parameters and chemical abundances of stars. Adding the stellar evolutionary models to the equation also enables the determination of the distances to the stars and consequently to the study of the motion of stars and their chemical composition in different layers of the Milky Way. The DR5 processing pipeline is optimized to handle a wide range of different normal star spectra in the observed sample, but it might not deliver optimal results for a subset of objects with more peculiar spectra \citep[e.g., spectroscopic binaries, active stars, cool giants, and metal-poor stars, ][]{2010AJ....140..184M, 2011AJ....141..200M, 2013ApJ...776..127Z}. For this reason, these types of objects need to be studied separately. Very metal-poor stars are particularly problematic because their spectra are
so featureless. As the central parts of the lines are excluded during the pipeline processing, there is little information left in the spectrum for the pipeline to rely upon. Consequently, the derived parameters, including the metallicity, are (sometimes strongly) affected (see Sect. \ref{sect_prop} for details on how the metallicity estimates are affected).

The first step in the analysis is to isolate the sample of candidate metal-poor stars. \citet{2010ApJ...724L.104F} used a criterion based on the metallicity and effective temperature derived in Data Release 2 \citep{2008AJ....136..421Z} to select the very metal-poor candidates for further analysis. Such a selection can be made very quickly since it uses readily available information. A drawback lies in the fact that some of the stars for which the pipeline results might have overestimated the metallicity could be excluded from the candidate list. Similarly, metal-richer stars with underestimated metallicities can contaminate the candidate sample. In order to avoid having to rely on a less reliable parameter-based selection, we opted to use the available spectra as the only source of information based on which we isolated the very metal-poor candidates.

One of the approaches that are suitable to isolate the metal-poor sample from the whole RAVE database is the projection of the spectra from their original wavelength space to a low-dimensional space where similar spectra are placed closely together. Afterward, a swift visual inspection is enough to find a group of interest that can be extracted by setting a bounding shape surrounding the region in the low-dimensional space. One such dimensionality reduction method (locally linear embedding) was used for classification purposes of RAVE spectra by \citet{2012ApJS..200...14M}. For the purpose of selecting the metal-poor stars, we chose to use the method called stochastic neighbor embedding \citep[SNE,][]{2002anips...15..833H,2008jmlr...9.2579M}. Various tests on different data sets proved that this method very efficiently projects complex data sets onto a plane while retaining  groups of similar data points. We refer to the original papers for a more thorough review of the method and only give a brief description here.

First, we resampled all radial-velocity-corrected and normalized spectra in the RAVE database to a common wavelength range between $8450\,\mathrm{\AA}$ and $8750\,\mathrm{\AA}$ in $D=768$ equally spaced wavelength bins. This step needs to be performed because spectra are not sampled at the same wavelength points. The wavelength range was slightly reduced from the original (about $30\,\mathrm{\AA}$ on each side) to avoid having to consider missing information in spectra with a large radial velocity correction. The number of bins were chosen so that the sampling remains roughly equal to the original and unnecessary over- or undersampling is avoided. Spectra with $S/N$ values below 10 were excluded, which reduced the sample size by about 20,000 stars. They do not carry enough information for subsequent analysis and make the low-dimensional projection less transparent, therefore they do not need to be kept in the dataset. In this way, we produced a data matrix whose rows represent the individual spectra. Alternatively, each spectrum $\mathbf{f}_i$ can now be viewed as a point in the $D$-dimensional space. We wish to evaluate how similar these points are to each other.

For every point in the $D$-dimensional space we define a Gaussian probability distribution centered on each point with a variance $\sigma^2_i$. The similarity between this point and the $j$-th point in the set is formalized through the conditional probability $p_{j|i}$ that $\mathbf{f}_i$ would pick $\mathbf{f}_j$ as its neighbor if neighbors are picked according to the probability given by the Gaussian distribution,
\begin{equation}
p_{j|i}=\frac{\exp\left( -\| \mathbf{f}_i - \mathbf{f}_j \|^2\right) / 2\sigma_i^2}{\sum_{k\neq i}\exp \left(-\| \mathbf{f}_i - \mathbf{f}_k \|^2\right) / 2\sigma_i^2}.
\label{eq_condprob}
\end{equation}
Throughout the text we use bold lower case symbols to denote vectors (i.e., $\mathbf{f}=\{f_i\}$) and bold upper case letters for matrices (i.e., $\mathbf{C}$). In a low-dimensional space, a similar expression can be written for the projected spectra $\mathbf{y}_i$ and $\mathbf{y}_j$,
\begin{equation}
q_{j|i}=\frac{\exp \left(-\| \mathbf{y}_i - \mathbf{y}_j \|^2\right)}{\sum_{k\neq i}\exp \left(-\| \mathbf{y}_i - \mathbf{y}_k \|^2\right)}.
\end{equation}
The variance in the last expression is set to $1/2$ for simplicity. This only affects the scaling of the final projection and therefore is of little importance. The probabilities $p_{j|i}$ and $q_{j|i}$ are equal when $\mathbf{y}_i$ and $\mathbf{y}_j$ correctly model their high-dimensional equivalents. In practice, this never occurs. To approach this condition as closely as possible, we require that the difference between the distribution is as small as possible. This is achieved by minimizing the sum of the Kullback-Leibler divergence \citep{Kullback59} over all data points. The cost function that we minimize is given by
\begin{equation}
C=\sum_i\sum_jp_{j|i}\log\frac{p_{j|i}}{q_{j|i}}.
\end{equation}
For the computations in this work we used an improved version of the method called t-distributed stochastic neighbor embedding (t-SNE) with the Barnes-Hut algorithm \citep{1986Natur.324..446B} used for faster cost function gradient approximation. The implementation we used was provided by \citet{2014jmlr...15.3221M}. There are two free hyper-parameters (perplexity and $\theta$) that govern how the projection is generated. Perplexity controls the variance of the Gaussian distribution in Eq. \ref{eq_condprob} and effectively determines the breadth of the region inside which the neighbors are sought (equivalently, the number of neighbors that a given spectrum is compared to). The second parameter $\theta$ ranges from 0 to 1 and is used to control the speed of the Barnes-Hut algorithm. Greater values allow the algorithm to operate faster, but it also becomes less accurate. When set to 0, the gradient is computed naively without approximations. In the final projection we set the value of perplexity to 50 and $\theta$ to 0.5. This enabled the calculation of the 2D projection of $\sim 420,000$ spectra within a day on a single CPU core. We also computed a 3D projection, but it did not provide any benefits over the 2D projection for the selection purposes.

\begin{figure}
\centering
\resizebox{\hsize}{!}{\includegraphics{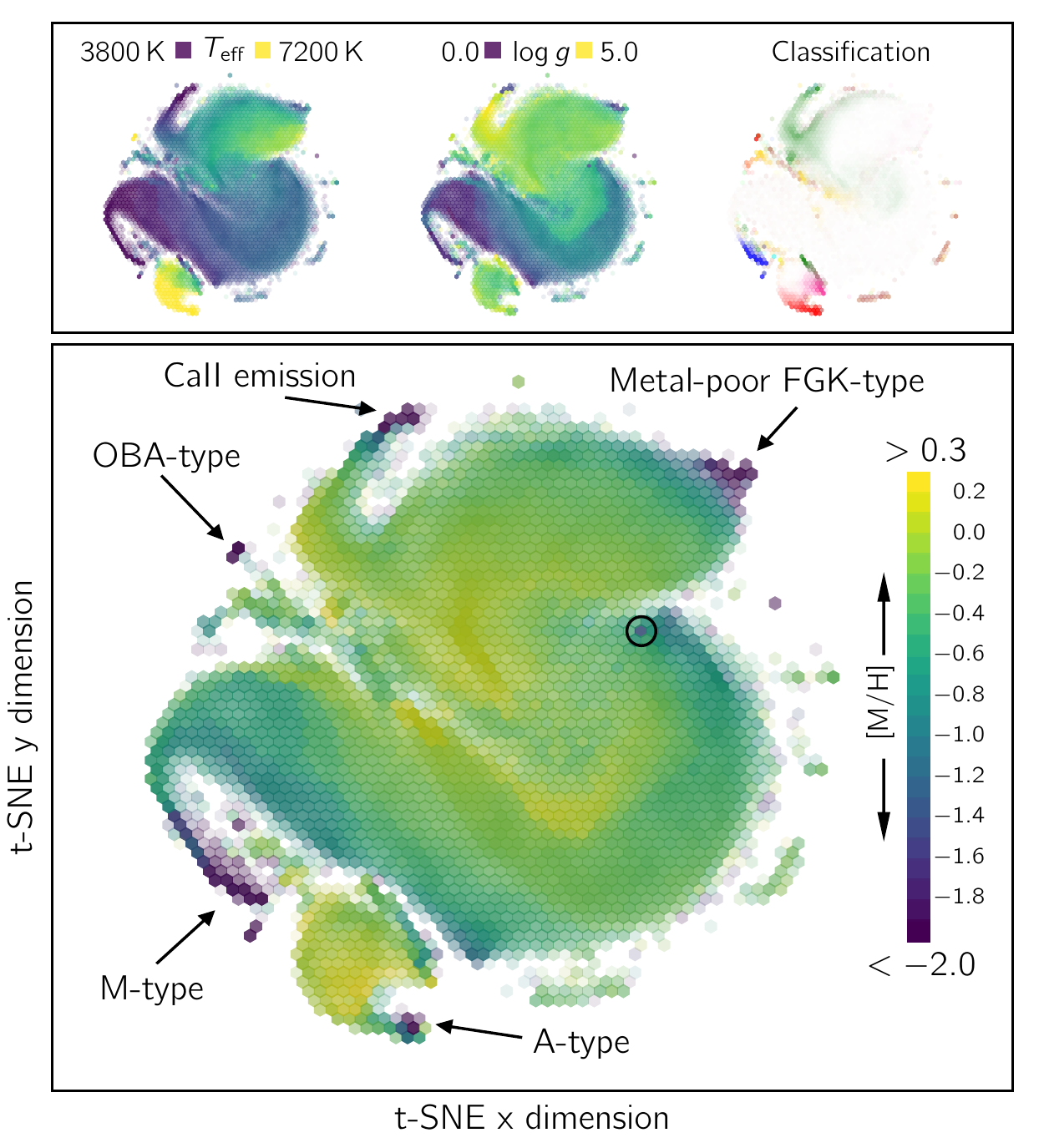}}
\caption{Bottom diagram: t-SNE projection computed from all available RAVE spectra with $S/N>10$. The individual points are binned into hexagons for easier visualization. The color scale encodes the median DR5 metallicity in each hexagon. The labels mark five groups for which the parameter estimation pipeline gives very low metallicities. The circled region also includes a few K-type metal-poor candidates that were included in the analysis. Top three diagrams: similar, but effective temperature, gravity and classification from \citet{2012ApJS..200...14M} are overplotted. The colors in the upper right diagram indicate different classes of peculiar spectra, while the white region contains normal single stars.}
\label{fig_tsne}
\end{figure}

The final projection with overplotted parameters is shown in Fig. \ref{fig_tsne}. Different regions of the projection correspond to stars with similar atmospheric parameters. The top part of the manifold includes mostly warmer dwarfs. The bottom part is mostly populated with cooler giants, with the exception of the island at the bottom, where A-type main-sequence stars reside. The outskirts of the projection, including the peninsula and islands, are occupied by peculiar classes of stars, while the main manifold includes the majority of the spectra ($\sim 90\%$) for which the parameters can be reliably derived using the main pipeline (i.e., normal stars). There are five distinct regions where the median metallicity from DR5 is very low. Inspection of the underlying spectra reveals that only one of these regions hosts spectra of metal-poor stars, while the others are populated by less common types of spectra that the pipeline confuses with metal-poor stars. The region that includes mostly very hot stars (OBA-type) also includes some peculiar spectra with strong emission lines as well as a few spectra with very shallow $\caii$ lines. The circled region in Fig. \ref{fig_tsne} also includes a handful of metal-poor candidates. We included the latter two groups along with the marked upper right region (FGK-type in Fig. \ref{fig_tsne}) in our metal-poor candidate sample that we analyzed further. Altogether,  $\nmpcandidate$ spectra were selected as our metal-poor candidates. The remaining three peculiar classes include M-type giants with temperatures below the limit of the synthetic spectral grid used by the pipeline ($3000\,\mathrm{K}$). Similarly, the temperatures of the A-type group stars are above the upper temperature limit of the grid ($8000\,\mathrm{K}$). The last group consists of spectra of stars that show signs of chromospheric activity \citep{2013ApJ...776..127Z}. A common feature in these spectra are elevated cores of the $\caii$ lines, which effectively makes them shallower than their non-active counterparts. For this reason, they are confused with metal-poor stars by the RAVE DR5 pipeline.

\section{Line profile modeling}
\label{sect_gpmodel}

Determining the elemental abundances requires measuring the equivalent widths (EWs) of the spectral lines. Typical very metal-poor star RAVE spectra with modest resolving power and $S/N$  have hardly any features that are distinguishable from the noise except for the near-infrared calcium triplet lines (Fig. \ref{fig_mp_spectra}). Any abundance analysis has to be based on these lines. These lines were recognized as a very good metallicity indicator decades ago by \citet{1988AJ.....96...92A}. Many researchers have adopted and extended this approach in the following years. \citet{2010A&A...513A..34S} and \citet{2013MNRAS.434.1681C} showed that the relation between the calcium abundance and metallicity performs well even for extremely metal-poor stars. According to \citet{2016MNRAS.455..199D}, the relation has a small spread for very metal-poor stars.

Initially, investigators obtained EWs by simply numerically integrating observed spectral lines over a predefined spectral window. The major drawback of this procedure is that any other signal that is present in the spectral range will also contribute to the EW measurement. To avoid this, Gaussian profiles were fit to the $\caii$ triplet lines and the best-fit synthetic profiles were integrated instead. Other functions such as Lorentzian, Moffat, or a combination of a Lorentzian and Gaussian profiles were also used to account for more complex line shapes in spectra of certain types of stars \citep[e.g.,][]{1997PASP..109..883R, 2004MNRAS.347..367C, 2006AJ....132.1630G, 2007AJ....134.1298C, 2015A&A...580A.121V}.

The most common procedure undertaken to fit the chosen profile to the spectral line in the literature is to define the cost function and minimize it using one of the robust solvers such as a Levenberg-Marquardt algorithm. Often, a standard $\chi^2$ statistic is used as a measure of the difference between the data and the model. This is easily justified if we assume that the noise in the data is independent (white, uncorrelated) and is drawn from a single distribution (i.e., a Gaussian with some variance). In this case, the logarithm of the likelihood of the parameters can be written as
\begin{equation}
\ln p(\mathbf{f}|\mathbf{x},\bm{\sigma},\bm{\theta})=-\frac{1}{2}\sum_{i=1}^N\left [ \frac{(f_i - m^{\bm{\theta}}(x_i))^2}{\sigma_i^2} + \ln(2\pi\sigma_i^2)\right ]
\label{eq_chi2}
.\end{equation}
We have denoted the observed flux measurements with $f_i$ measured at $N$ wavelength points $x_i$. The uncertainties of the measurements are $\sigma_i$. Individual model fluxes are computed where the data are sampled. The model $m^{\bm{\theta}}$ parameters are denoted by $\bm{\theta}$. In case of constant $\sigma_i$, Eq. \ref{eq_chi2} reduces to the standard definition of $\chi^2$. However, this expression is only valid if the assumptions we made about the noise are not violated. When a profile is fit
to the spectral lines, this often does not hold. Three major contributors to the correlations in the noise are i) fringing, ii) the residual signal left in the data during the reduction phase since the normalization procedure is not perfect, and iii) weaker and blended spectral lines that lie in the vicinity of the spectral lines of interest. The first effect is inevitably introduced when the spectrum is normalized. Some sort of polynomial function is typcially used to model the continuum of the spectrum. As this never perfectly matches the real continuum, a slowly oscillating signal remains in the normalized spectrum. The second contribution can also be rarely avoided. The weakest lines might not be resolved, but they can still induce correlations between adjacent flux points, making the noise non-white. The same is
true for the weak lines that are blended with a much wider profile of the line of interest. Failing to take these effects into consideration when fitting a profile can lead to biased estimates of the parameters as well as underestimated uncertainties, the latter aspect being particularly misleading. Although we might not be specifically interested in what caused the correlations or what the values of the parameters that describe them are, it is vital to have enough flexibility in the model to be able to account for them.

\subsection{Noise model for Gaussian processes}
\label{sec_gpmodel}

One of the possibilities to add an extra layer of flexibility to model the noise in the spectra is to use a Gaussian processes framework \citep{Rasmussen:2005:GPM:1162254}. They have been successfully applied to many fields in astronomy in the recent years. \citet{2009ApJ...706..623W} used it to model SDSS galaxy redshifts and \citet{2012MNRAS.419.2683G} and \citet{2013ApJ...772L..16E} modeled instrumental systematics in transmission spectroscopy with the help of Gaussian processes. This approach gained significant popularity in modeling the residual signal in time series analysis. Several authors used them for studies of exoplanets and stellar variability \citep[e.g.,][]{2014ApJ...791...89D,2015MNRAS.447.2880A,2015ApJ...800...46B,2015MNRAS.452.2269R}. \citet{2015ApJ...812..128C} used Gaussian processes to account for the mismatches between the observed and synthetic stellar spectra. We only give a short derivation of the formalism regarding Gaussian process. For a more thorough description we refer to \citet{Rasmussen:2005:GPM:1162254}.

We start by rewriting Eq. \ref{eq_chi2} in a more general form using matrix notation,
\begin{equation}
\ln p(\mathbf{f}|\mathbf{x},\bm{\sigma},\bm{\theta})=-\frac{1}{2}\mathbf{r}^T \mathbf{C}^{-1} \mathbf{r} - \frac{1}{2}\ln|\mathbf{C}| - \frac{N}{2}\ln(2\pi),
\label{eq_likelihood}
\end{equation}
where $\mathbf{C}=\bm{\sigma}\mathbf{I}$ is a diagonal covariance matrix with $\sigma_i^2$s along the diagonal and $\mathbf{r}=\mathbf{f}-\mathbf{m}_{\bm{\theta}}$. To account for more complex correlations between flux points, we can expand the covariance matrix with an additional kernel $\mathbf{K}_{\bm{\alpha}}$ by adding it to the diagonal covariance matrix,
\begin{equation}
\mathbf{C}=\bm{\sigma}\mathbf{I} + \mathbf{K}_{\bm{\alpha}}.
\end{equation}
The parameters $\bm{\alpha}$ describe the noise model. There are a number of kernels to choose from. The squared exponential kernel, where the correlations between the individual points ($i,j$-th element of the matrix) are expressed as
\begin{equation}
k_{\bm{\alpha}}(x_i, x_j)=A^2\exp\left(-\frac{(x_i-x_j)^2}{2w^2} \right),
\label{eq_sqexp}
\end{equation}
is a simple but convenient choice for modeling a smooth correlation between data points. Other kernels are better suited when modeling less smooth variations or periodic signals, for example. \citet{Rasmussen:2005:GPM:1162254} give an overview of the possible choices. In the last equation, the two kernel parameters $\bm{\alpha}$ are the amplitude of the variations $A$ and their characteristic length-scale $w$. The higher the value of the latter parameter, the longer the correlations that are described by this kernel. After defining the kernel, we can generate zero mean random functions $\mathbf{y}_*$ over a selected domain $\mathbf{x}_*$ by drawing from the normal distribution
\begin{equation}
\mathbf{y}_*\sim\mathcal{N}(\mathbf{0},\mathbf{K}_{\bm{\alpha}}(\mathbf{x}_*, \mathbf{x}_*)).
\label{eq_random_function}
\end{equation}
Different kernels can be easily added together to produce more complex covariance functions when necessary.

\begin{figure}
\centering
\resizebox{\hsize}{!}{\includegraphics{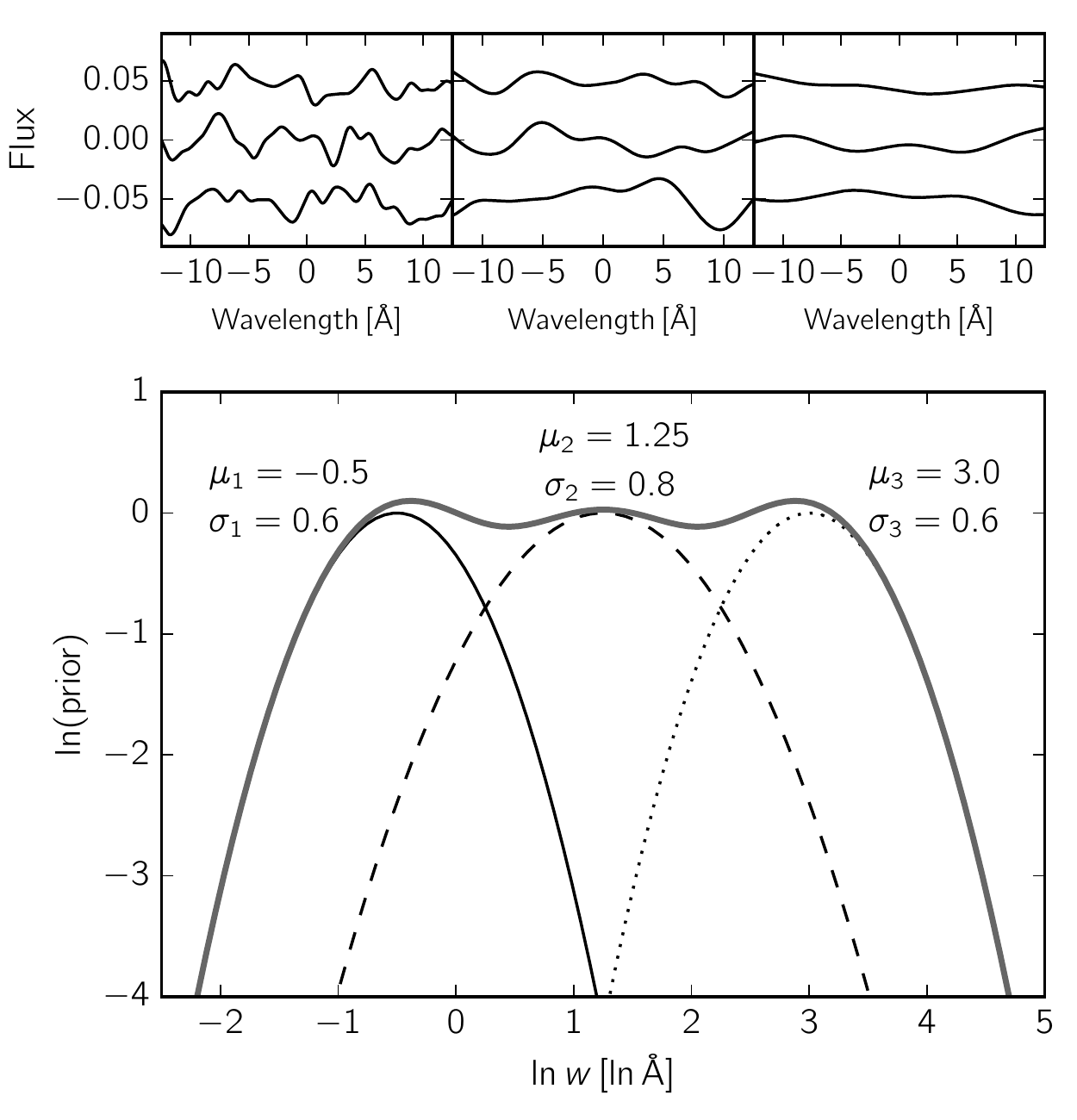}}
\caption{Bottom panel: priors imposed on the length scale of the three squared exponential kernels used for the profile noise modeling. The labels $\mu$ and $\sigma$ denote the mean and the standard deviation of each Gaussian distribution, respectively. The thick gray line is the sum of all three. Each of the top three panels shows three functions generated by each kernel and $w$ values randomly drawn from the corresponding prior. The amplitudes were chosen so that they roughly match the typical variations in the RAVE spectra. The functions are separated by $0.05$ in the vertical direction for clarity.}
\label{fig_w_priors}
\end{figure}

A more powerful aspect of Gaussian processes is their ability to make predictions given the data and their uncertainties. For a wavelength point $x_*$ somewhere inside of the wavelength range $\mathbf{x,}$ it can be shown that the average predicted flux at this point is equal to
\begin{equation}
\bar{f_*}=\mathbf{k}_*^T\mathbf{C}^{-1}\mathbf{f},
\label{eq_flux_mean}
\end{equation}
where $\mathbf{k}_*=\mathbf{k}(x_*)$ is a vector of covariances between the test point $x_*$ and the wavelengths of the observed spectrum. The variance (or square of the uncertainty) of this newly predicted point is
\begin{equation}
\mathrm{Var}(f_*)=k(x_*,x_*)-\mathbf{k}_*^T\mathbf{C}^{-1}\mathbf{k}_*.
\label{eq_flux_variance}
\end{equation}
The first term on the right side of the equation is simply $A^2$ for the case of the squared exponential kernel. With this feature we can make a prediction around the mean profile in such a way that the Gaussian processes component takes over the deviations from the mean profile and effectively describes the noise.

There are number of available software packages that efficiently implement Gaussian processes related computations. For the purpose of modeling the noise in the $\caii$ line profiles, we used the implementation provided by \citet{10.1109/TPAMI.2015.2448083}.

\subsection{Complete model}

To describe the shapes of all three $\caii$ lines, we adopted a Voigt profile (a convolution of a Lorentzian profile with a Gaussian). \citet{1997PASP..109..883R} noted that in the metal-poor regime, these lines are well described by a Gaussian function, but to give more freedom to the profile in case there is a need to model possibly more extended wings of the lines, the Voigt profile was chosen. A Voigt profile has three parameters, the amplitude $\beta$, the scale parameter of the Lorentzian profile $\gamma$, and the standard deviation of the Gaussian $\Sigma$. We modeled each of the three lines indexed by $l$ with its own Voigt profile, so that
\begin{equation}
\mathbf{m}_l=\mathrm{Voigt}(\mathbf{x}_l;\beta_l, \gamma_l, \Sigma_l,\lambda_0).
\end{equation}
We also allowed the centers of the profiles to vary in wavelength by $\lambda_0$. This parameter was shared by all three profiles, because possible radial velocity offsets were assumed to be equal for all three lines. The vector $\mathbf{x}_l$ denotes the spectral range of each line. Altogether, we have ten profile parameters $\bm{\theta}=(\bm{\theta}_1,\bm{\theta}_2,\bm{\theta}_3,\lambda_0)$, three for each of the line profiles, and one for the common wavelength shift.

The choice of the kernel used in the noise modeling is arbitrary as there is no prescription of which one is best used in a given situation. We used three separate squared exponential kernels added together, each for a different variation length-scale. We constrained them to a selected range by imposing three prior distributions on the parameters $w$ of each of the three kernels. This enabled us to have a relatively flat but still smooth probability over a broader range of variation scales. The limits were chosen so that on the short-scale end the prior probability becomes very low for variation lengths comparable to the distance between pixels. On the longer side, we cut off the scale length at about twice the size of the wavelength window that is used for the profiles ($\sim 50\,\mathrm{\AA}$). In this way, we allow for almost constant offsets (almost non-varying functions) within the window. The priors are shown in Fig. \ref{fig_w_priors}. For illustration, we also plot the functions generated by each of the kernels according to Eq. \ref{eq_random_function} with length-scale parameters $w_\varkappa$ for individual kernels $\varkappa$ randomly drawn from the corresponding prior. We assumed that the correlations in the noise are similar across the ranges of all three lines, so we only use a single three-kernel model for all three.

To combine this, we write down the total likelihood for all three profiles $l$ as
\begin{equation}
\ln p (\mathbf{f}|\mathbf{x},\bm{\sigma},\bm{\theta},\bm{\alpha})= \sum_{l=1}^3 \ln p_l(\mathbf{f}_l|\mathbf{x}_l,\sigma_l,\bm{\theta}_l,\bm{\alpha}),
\label{eq_total_likelihood}
\end{equation}
where the individual likelihoods are given in Eq. \ref{eq_likelihood} and now also include the kernel parameters $\bm{\alpha}=(\bm{\alpha}_1,\bm{\alpha}_2,\bm{\alpha}_3)$. Each of the $\bm{\alpha}_\varkappa$ has two components, the amplitude and the length scale. The components of the three covariance matrices that enter the upper equation through the individual likelihoods are equal to
\begin{equation}
c_{ij,l}=\sum_{\varkappa=1}^3 k_{\alpha_\varkappa}(x_{\varkappa,i},x_{\varkappa,j}) + \sigma_l\delta_{ij}.
\end{equation}
The missing index $l$ next to the Gaussian processes kernel indicates that the kernel does not change among the profiles, but we allow the white-noise component $\sigma_l$ to vary among them. We also assume that the white-noise component is equal for all the points within the range of a single profile as we have no information on individual flux point uncertainties. We approximate its value as $S/N^{-1}$ within the profile range. The symbol $\delta_{ij}$ is the standard Kronecker delta.

\begin{figure*}
\centering
\includegraphics[width=17cm]{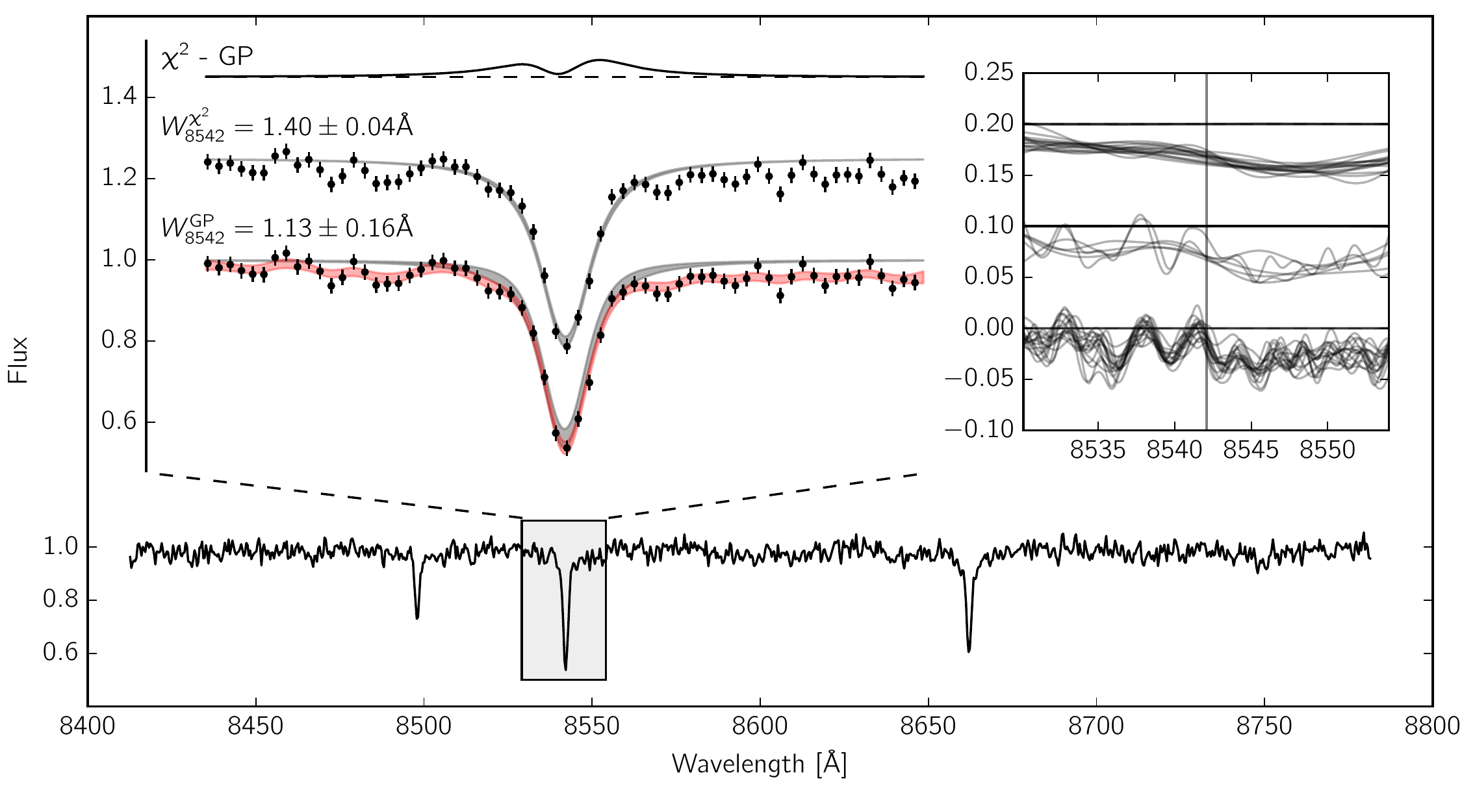}
\caption{Bottom part: one extremely metal-poor star spectra of
RAVE. The metallicity of this star was estimated by our study to be $\met = -3.04\pm 0.22\,\mathrm{dex}$ and the measured S/N per pixel is $\sim 52$. The gray box marks the region around the central $\caii$ line for which the two solutions and their difference are plotted above. The gray shaded areas give $\pm 1\sigma$ Voigt profile solutions for the regular $\chi^2$ approach (top) and the noise-modeled approach (bottom) with measured equivalent widths of the lines given next to each solution. In the latter case there is also a full model $\pm 1 \sigma$ region shown in red (see text for details). The inset shows 20 draws from each of the three Gaussian process kernel posteriors using Eq. \ref{eq_flux_mean}. The second and third group are vertically shifted by 0.1 and 0.2 for clarity.}
\label{fig_mp_spectrum}
\end{figure*}

The likelihood in Eq. \ref{eq_total_likelihood} can be maximized to obtain the best-fitting values for all 16 parameters in the model (ten for the profiles and six for the kernels). However, we can also compute the posterior distributions of all the parameters by imposing prior distributions on the rest of the parameters in addition to the length-scales. The posterior follows from the Bayes rule,
\begin{equation}
\ln p(\bm{\theta},\bm{\alpha}|\mathbf{f},\mathbf{x},\bm{\sigma}) = \ln p(\mathbf{f}|\mathbf{x},\bm{\sigma},\bm{\theta},\bm{\alpha}) + \ln p(\bm{\theta},\bm{\alpha}) - \ln Z.
\label{eq_posterior}
\end{equation}
The last term on the right-hand side is an unimportant normalization constant for our purposes and only scales the posterior distribution.

\subsection{Sampling the posterior distribution}

Using Eq. \ref{eq_posterior}, we can generate the posterior distribution by sampling it with a Markov chain Monte Carlo (MCMC) sampler, for example. For this study we used a popular implementation of an affine-invariant ensemble sampler called \texttt{emcee} \citep{2010camcs....5.65G,2013PASP..125..306F}. The procedure was as follows. Each of the $\nmpcandidate$ metal-poor candidate spectra was analyzed separately. First, we isolated the $\caii$ lines by selecting $\pm 12\,\mathrm{\AA}$ regions from the center of each line $(8498.02\,\mathrm{\AA}, 8542.09\,\mathrm{\AA}, 8662.14\,\mathrm{\AA})$, which resulted in each line and its surroundings being sampled at 64 wavelength points. The uncertainty for all measured flux data points was assumed to be the same and was computed as the inverse of the $S/N$ derived by the DR5 pipeline. We initialized the positions of the \texttt{emcee} walkers in a small Gaussian ball centered on the average values for length scales and profile parameters obtained from a few test runs. We tested how varying the initial locations influenced the final distribution and found that within reasonable limits no changes are observed. Amplitudes of the Gaussian process kernels were drawn from a uniform distribution across the prior space. We have already specified the prior distributions for length
scales. For amplitude priors we chose flat distributions in natural logarithm between $-50$ and $50$. Wavelength offsets were constrained with a uniform prior between $-5\,\mathrm{\AA}$ and $5\,\mathrm{\AA}$. The remaining profile parameters $(\beta_l, \gamma_l, \Sigma_l)$ were required to be greater than zero.

After running the sampler for 200 iterations (per ensemble member), we restarted the sampling with initial locations of the members around a Gaussian ball centered on the highest likelihood point thus far. We ran the sampling for another 700 iterations. After the first set of samples was obtained, we recalculated the $S/N$  to obtain a better estimate within each profile range. This was achieved by generating the prediction of the long and medium length-scale noise component using Eq. \ref{eq_flux_mean} with appropriate kernels $(\varkappa=2,3)$. After subtracting the average profile and the generated correlated noise from the data, we calculated the standard deviation of the residuals and treated this as our new $S/N$ value. This is of course not a perfect estimate since there are still some short-scale correlated variations present in the remainder, but it is a better approximation than our initial value. From this point on, each of the three line regions has its own $S/N$ estimate. We repeated these steps by first running the sampler using the new $S/N$ for 1000 iterations, repositioning the walkers, and finally running it for 6000 iterations. The whole procedure is computationally quite intensive as it involves the calculation of the inverses of three covariance matrices with each iteration. Obtaining the posterior distribution for a single star takes around half  a CPU hour.

Equivalent widths were computed by randomly selecting 3000 points from the last 5000 iterations and generating each of the three profiles individually. We note that only profile models were generated without noise contribution, but because we are selecting the solutions randomly from the posterior distribution, this also ensures that we are correctly marginalizing over the noise model parameters. We integrated the profiles over a broad region ($\pm \infty,$ but in practice, this was done over $\pm 75\,\mathrm{\AA}$) to ensure that the signal from possibly extended tails is taken into account. This was performed separately for each individual line. It ensures that even when the wings of the profile are very far reaching, the whole contribution is accounted for in the EW measurement. Final EW estimates were calculated as means of the posterior distributions of individual EWs and their uncertainties as standards deviations of these distributions.

One of the great problems in using a MCMC sampler to sample from the posterior probability is knowing how many iterations are required in order for the sampler to a reach stationary state. We analyzed the convergence of the EWs for several cases. The EW chains settle down after about 500 iterations starting from the initial point, and subsequent variations are well within the uncertainty and can therefore be neglected.

The results for one of the profiles are shown in Fig. \ref{fig_mp_spectrum}. The diagram shows a spectrum of an extremely metal-poor star with a $S/N$ typical for our sample. For clarity, the resulting profile of only the strongest line is shown. The red band shows a $\pm 1\sigma$ deviation from the complete model. The gray band shows the same, but only for the profile model (i.e., the difference between the complete model and the noise). It is evident that there are noticeable systematic differences between these two, meaning that noise contributions are highly non-Gaussian. The complete model traces the continuum and local variations very well. In turn, the profile model in this case is consequently slightly shallower. To assess how the solution behaves in absence of the noise model, we fit the lines of the same spectrum with a simple $\chi^2$ noise model by optimizing Eq. \ref{eq_chi2}. The results are shown in the same diagram. While the profiles might not look very different, their equivalent widths reveal how much influence the more complex noise model has. The top profile has a significantly larger EW, while its uncertainty is three times smaller than in the bottom case. To illustrate how strong the noise model contribution from the three different kernels is, we plotted 20 random draws from the posterior and only generated a partial noise model from each kernel.

\begin{figure}
\centering
\resizebox{\hsize}{!}{\includegraphics{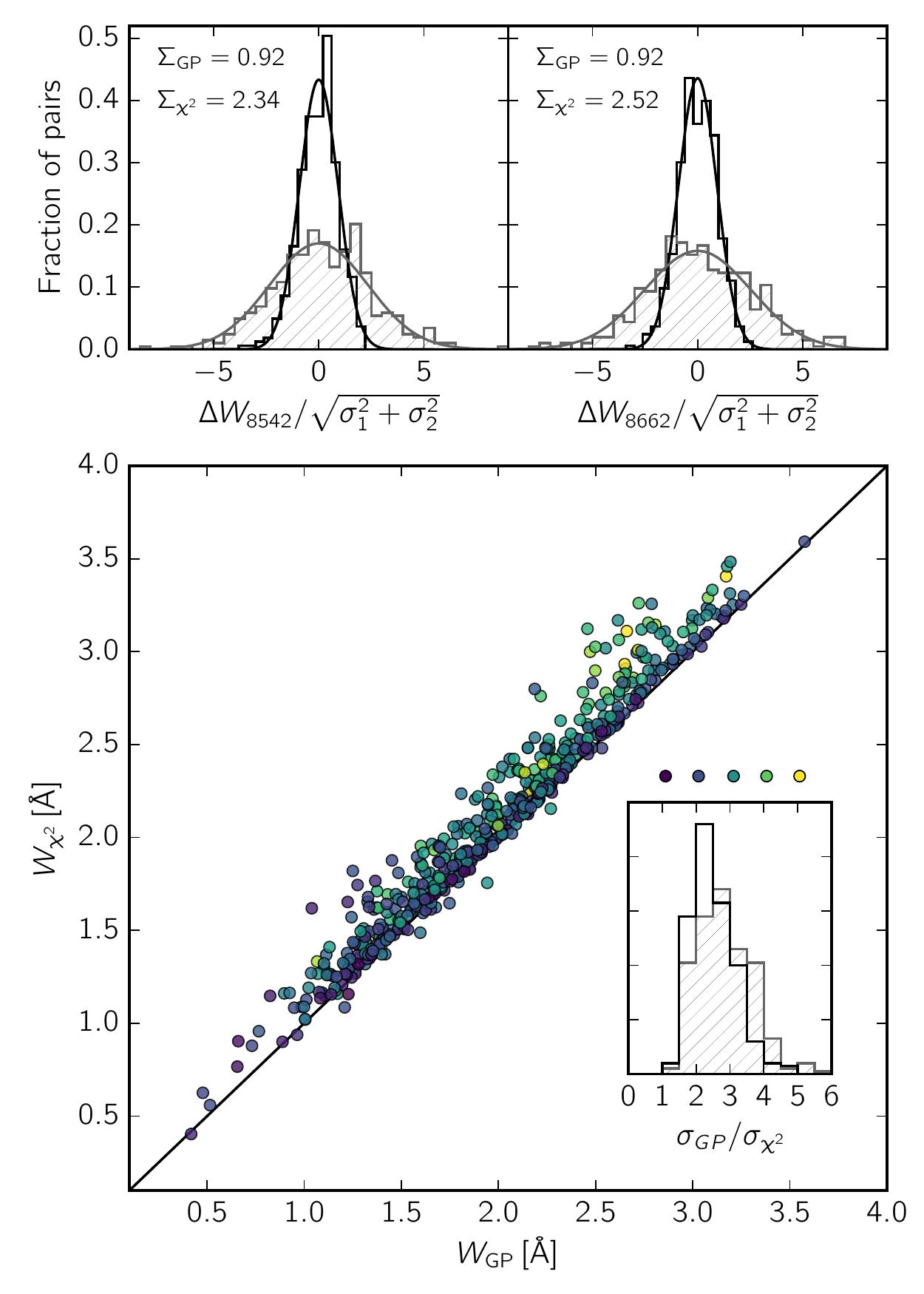}}
\caption{Bottom diagram: comparison between the EWs measured using the Gaussian processes (GP) and EWs coming from a simple $\chi^2$ analysis. EW measurements for the line at $8498\,\mathrm{\AA}$ are not shown. The inset shows the histogram of the uncertainty ratios. Brighter colors of the dots correspond to higher ratio values. Top diagrams: histograms of the differences between the $i$-th and the $j$-th repeated observation for the lines at $8542\,\mathrm{\AA}$ and $8662\,\mathrm{\AA}$. The black histogram includes the measurements generated by Gaussian processes, the gray histogram shows the $\chi^2$ derived values. Gaussian distributions with equivalent standard deviations $\Sigma$ are overplotted, which should ideally be equal to 1.}
\label{fig_gp_chi2}
\end{figure}

To further explore how the addition of the noise model influences the EW measurements, we computed the $\chi^2$ solutions for a subset of 153 stars for which we have at least one repeated observation and compared them to the complete model solutions. The results are shown in Fig. \ref{fig_gp_chi2}. In most cases, the $\chi^2$ measurements are overestimated compared to the Gaussian process values. Even more importantly, the uncertainties for $\chi^2$ solutions are almost exclusively smaller than their counterparts derived from Gaussian processes. We estimated how realistic the uncertainties are for both models by computing the differences between the $i$-th and the $j$-th repeated observation (computed for all possible pairs where three or more observations were available) of the same star in units of the total uncertainty. For the $\chi^2$ model the results show that the uncertainties are heavily underestimated, while the Gaussian process model overestimates the uncertainties by only a few percent, as shown in Fig. \ref{fig_gp_chi2}.

\section{Metallicity calibration}
\label{sect_met}

A method of converting the equivalent widths of $\caii$ lines into metallicity\footnote{We use the label metallicity for $\met$ through this text.} estimates that became popular and was used in many studies was first introduced by \citet{1991AJ....101.1329A} and \citet{1991AJ....101..515O}. Studying the stars in Galactic and LMC clusters, they realized that the $\met$ and the difference between the absolute magnitude of a star and the horizontal branch of a cluster are roughly
linearly related, offset by the sum of the equivalent widths of the two strongest $\caii$ lines (so-called reduced equivalent width). Other investigators followed the same idea and derived empirical relations for globular clusters \citep[e.g.,][]{1997PASP..109..883R} and dwarf spheroidal galaxies \citep[e.g.,][]{2001MNRAS.327..918T,2008MNRAS.383..183B}. \citet{2004MNRAS.347..367C} and \citet{2004oee..sympE...8C} extended the calibration over a wider range of ages, and \citet{2010A&A...513A..34S} and \citet{2013MNRAS.434.1681C} derived the relation for very metal-poor stars. \citet{2016MNRAS.455..199D} showed that the calcium near-infrared triplet measurements are a reliable metallicity estimator and that basing metallicity estimates on the $\caii$ triplet does not introduce significant biases. \citet{2015A&A...580A.121V} added a quadratic term in reduced equivalent width to the relation to better describe the relation of K-type giants in the Galactic bulge. \citet{2007AJ....134.1298C} introduced similar relations, but used other luminosity indicators such as absolute $I$ and $V$ magnitudes in their relations. These relations work well for stars at known distances, for which we can calculate the absolute magnitudes, but cannot be applied to field stars at unknown distances. The RAVE catalog does provide distances to the observed stars, but they were derived by relying on the parameters from the DR5 pipeline and are therefore potentially biased. In addition, stellar evolutionary models used to determine the distance only extend to $\met\sim -2.0$, which also makes distance estimates unreliable for very metal-poor stars past this limit.

An alternative approach was taken by \citet{2010ApJ...724L.104F} to obtain the abundances of the observed stars. They calculated the equivalent widths and processed them with the \texttt{MOOG} program \citep{1973ApJ...184..839S} to generate abundance estimates using the LTE one-dimensional Kurucz model atmospheres. While this approach does not require the knowledge of the distance to the star, it still relies on the derived temperatures and gravities that are used to select the appropriate model atmosphere. Any offsets in these parameters can lead to offsets in abundance measurements. In what follows, we describe a method that relies solely on the information in form of equivalent widths and stellar colors.

\subsection{Method}

In the following approach, we try to circumvent the aforementioned difficulties by only using reliable pieces of data in the form of equivalent widths and photometric measurements. Stars in the RAVE catalog were also observed by the AAVSO Photometric All-Sky Survey \citep[APASS,][]{2014AJ....148...81M} in four visual and one near-infrared band. Spectral ranges deeper in the infrared domain were covered by DENIS, 2MASS, and WISE. Based on the completeness criterion, we only used measurements from the 2MASS and WISE surveys (not all candidate stars have APASS magnitudes) to maximize the number of stars to which the calibration can be applied. In addition, the effect of reddening on the colors is substantially decreased further into the infrared spectral range. We only employed the EW measurements of the stronger two $\caii$ lines. The bluest line in some spectra is mixed in with an emission ghost that is caused by the zero-order reflection, which makes the EW measurements unreliable \citep{2008AJ....136..421Z}. Our set of $M=5$ \textit{\textup{observables}} that are used in the calibration relation is defined as follows:
\begin{equation}
\mathbf{o} = \left[W_{8542}, W_{8662}, J-H, H-K_S, w_1-w_2\right],
\label{eq_observables}
\end{equation}
and it consists of two EW measurements and three colors (without the extinction). The first two are from the 2MASS survey, while the last one comes from the WISE survey. The inclusion of infra-red colors is important since the equivalent widths themselves do not carry enough information to decouple the effective temperature and gravity degeneracy that can be clearly seen in Fig. \ref{fig_mp_spectra}. Our approach for decoding it was motivated by the flexible data-driven model \textit{The Cannon} introduced by \citet{2015ApJ...808...16N}. They used it to derive stellar atmospheric parameters from  spectra. Its core is represented by a linear model $\bm{\Theta} \cdot \bm{\zeta}$ through which the observed values (fluxes at a given wavelength) are connected to the labels ($T_\mathrm{eff}$, $\log g$, $\met$). The variable $\bm{\Theta}$ describes the coefficients of the model and $\bm{\zeta}$ gives the label vector, for which we have to choose the form. Our metallicity calibration can be constructed in a similar fashion when we replace the observed fluxes with the observables from Eq. \ref{eq_observables}. In that case, the likelihood of the parameters of the star $i$ and observable $m$ becomes
\begin{equation}
\ln p(o_{im}|\bm{\zeta}_i,\sigma_{im},\bm{\Theta}_m,s_m)=\frac{1}{2}\frac{\left (o_{im} - \bm{\Theta}_m \cdot \bm{\zeta}_i\right)^2}{s_m^2 + \sigma_{im}^2} - \frac{1}{2}\ln\left(s_m^2+\sigma_{im}^2\right).
\label{eq_cannon_lnlike}
\end{equation}
Uncertainties of the observables enter the upper likelihood through $\sigma_{im}^2$. We also allowed for the intrinsic scatter of the observables by introducing an additional noise term $s_m^2$. The simplest form of the label vector when the only label of interest is $\met$ is
\begin{equation}
\bm{\zeta}_i = \left[1, \met_i\right].
\end{equation}
If we wished for the model to be more ambitious and also provide temperatures, we could write the label vector as
\begin{equation}
\bm{\zeta}_i = \left[1, T_\mathrm{eff,i}, \met_i\right].
\end{equation}
We chose the simplest linear form for the labels. Of course, more complex forms are possible. For example, adding quadratic terms to the label vector will enable the model to learn about the correlations between the labels and therefore it will perform better. It will also add complexity, however, which will extend the computation time.

After the label vector is decided upon, the model can be trained. For this purpose, we also need a set of reference stars for which we have the observables and know their labels. The source of the training set is discussed in the next subsection. To train the model, we optimize for the coefficients $\bm{\Theta}$ and the intrinsic scatter parameter $s_m^2$ by finding the maximum of the sum of the likelihood for individual stars for each observable separately. When evaluating the observables of a star with unknown labels, we find the maximum of the same likelihood but using the optimal values of the coefficients and summing them over all observables. Here we optimize for the unknown labels.

There are two drawbacks in the described procedure. As already mentioned, the choice of the label vector must be made. If the label complexity is too small, the model will not be able to describe the encoded relations well; if it is too large, it will be slow and prone to overfitting. Another detail we have not included in the model so far are the uncertainties of the labels themselves. If they are not small, they should not be neglected. To address both issues, we introduced an upgraded model where instead of settling for a predefined set of label functions, we allow their form to be optimized along with the rest of the parameters so they are \textit{learned} from the data. A very convenient way of defining a pool of smooth functions are Gaussian processes. Here we are not using Gaussian processes to model the noise as was the case when modeling the spectral lines, but rather as a pool of functions that we use to describe the metallicity relation. The linear model in Eq. \ref{eq_cannon_lnlike} is replaced by the Gaussian process kernel (see \citet{Rasmussen:2005:GPM:1162254} for details of how a linear model relates to Gaussian processes). We write the generalized likelihood function for the $m$-th observable as
\begin{equation}
\ln p(\mathbf{o}_m|\bm{\Xi},\bm{\sigma}_{m},s_m,B_m,\bm{\omega}_m) = -\frac{1}{2}\mathbf{o}_m^T\mathbf{C}_m^{-1}\mathbf{o}_m-\frac{1}{2}\ln|\mathbf{C}_m|-\frac{n}{2}\ln 2\pi.
\label{eq_gpcannon_lnlike}
\end{equation}
We denoted the amplitude with $B_m$ and all single observables for all calibrating stars with $\mathbf{o}_m$. The variable $\bm{\Xi}$ is a $n\times D$ matrix where each of its rows $\bm{\xi}_i=[T_\mathrm{eff,i}, \met_i]$ contains labels for each of the $n$ calibration sample stars. Both the amplitude and the matrix along with the observable uncertainties $\bm{\sigma}_{m}$ and intrinsic scatter $s_m^2$ enter the equation through the covariance matrix, whose components are
\begin{align}
c_{ij,m} & = B_m^2\exp\left(-\frac{1}{2} \sum_{d=1}^3\frac{\left(\xi_{id} - \xi_{jd}\right)^2}{\omega_{md}^2} \right) + \left(s_m^2 + \sigma_{im}^2\right)\delta_{ij}\\\nonumber
             & = B_m^2\exp\left(-\left(\bm{\xi}_i - \bm{\xi}_j\right)^T\mathbf{W}_m^{-1}\left(\bm{\xi}_i - \bm{\xi}_j\right)\right) + \left(s_m^2 + \sigma_{im}^2\right)\delta_{ij}.
\label{eq_gpcannon_cij}
\end{align}
We used $\omega_{m}$s to introduce the length scales for each label. These parameters control how smooth the calibration relation
is with respect to each of the dimensions. In the last row we rewrote the expression using the vector notation and introduced
\begin{equation}
\mathbf{W}_m = \mathrm{diag}(\bm{\omega}_m^2/2).
\end{equation}
To account for the uncertainties of the labels, we follow \citet{dallaire2009}. Instead of using only point estimates, we assume that individual star labels are normally distributed around the mean values $\bar{\bm{\xi}}$ and have their own covariance matrices $\bm{\Sigma}$,
\begin{equation}
\bm{\xi}\sim \mathcal{N}\left(\bar{\bm{\xi}},\bm{\Sigma}\right).
\end{equation}
In this case, the components of the Gaussian process covariance matrix are expressed as
\begin{align}
c_{ij,m} & = B_m^2\frac{\exp\left(-\left(\bar{\bm{\xi}_i} - \bar{\bm{\xi}_j}\right)^T\left[\mathbf{W}_m+\bm{\Sigma}_i+\bm{\Sigma}_j\right]^{-1}\left(\bar{\bm{\xi}_i} - \bar{\bm{\xi}_j}\right)\right)}{\left |\mathbf{I} + \mathbf{W}_m^{-1}\left( \bm{\Sigma}_i+\bm{\Sigma}_j\right)\right |^{1/2}} \\\nonumber
 & + \left(s_m^2 + \sigma_{im}^2\right)\delta_{ij}.
\end{align}

With the likelihood function in place, we need to train the model using a calibration set. This is done by optimizing the kernel parameters $B_m$ and $W_m$, and the intrinsic scatter $s_m^2$ for each of the observables. For the sake of simplicity, we only use the best-fitting values of these parameters for further computations. To properly account for the fact that these parameters are distributed themselves, we should have marginalized over them. Since this means computing the inverse of the covariance matrix many more times, the problem becomes too computationally intensive. The tests we performed by marginalizing over a limited number of possible kernels revealed that using only best-fitting values differs negligibly from the marginalized case.

The evaluation of the set of observables $\mathbf{o}_*$ and their uncertainties $\bm{\sigma}_*$ for a test star with unknown labels is made in much the same way as discussed in Section \ref{sec_gpmodel}. With a label set provided by the optimizing algorithm at each iteration and using the best-fit values from the first step, we generate the expected value and its uncertainty for each of the observables using the equivalents of Eqs. \ref{eq_flux_mean} and \ref{eq_flux_variance},
\begin{equation}
\tilde{\mu}_{m}=\mathbf{k}_*^T\mathbf{C}_m^{-1}\mathbf{o}_m,
\label{eq_label_mean}
\end{equation}
\begin{equation}
\tilde{\sigma}_{m}=B_m^2-\mathbf{k}_*^T\mathbf{C}_m^{-1}\mathbf{k}_*.
\label{eq_label_cov}
\end{equation}
We use $\mathbf{k}_*$ to denote the covariances between the training label set and the current iteration test point labels. The likelihood we wish to optimize in order to obtain the best values for the labels is then
\begin{equation}
\ln p = \sum_{m=1}^M\left( \frac{1}{2} \frac{(o_{m,*}-\tilde{\mu}_m)^2}{\tilde{\sigma}_m^2 + \sigma_{m,*}^2} - \frac{1}{2} \ln \left( \tilde{\sigma}_m^2 + \sigma_{m,*}^2 \right)\right).
\label{eq_label_likelihood}
\end{equation}
We left out the normalizing factor that plays no role during optimization. The best-fitting value can be computed using one of the numerous optimization algorithms. If we require the posterior distribution, we can multiply the likelihood with a prior distribution over the labels and use an MCMC sampler to generate it.

\subsection{Calibration sample}

\begin{figure}
\centering
\resizebox{\hsize}{!}{\includegraphics{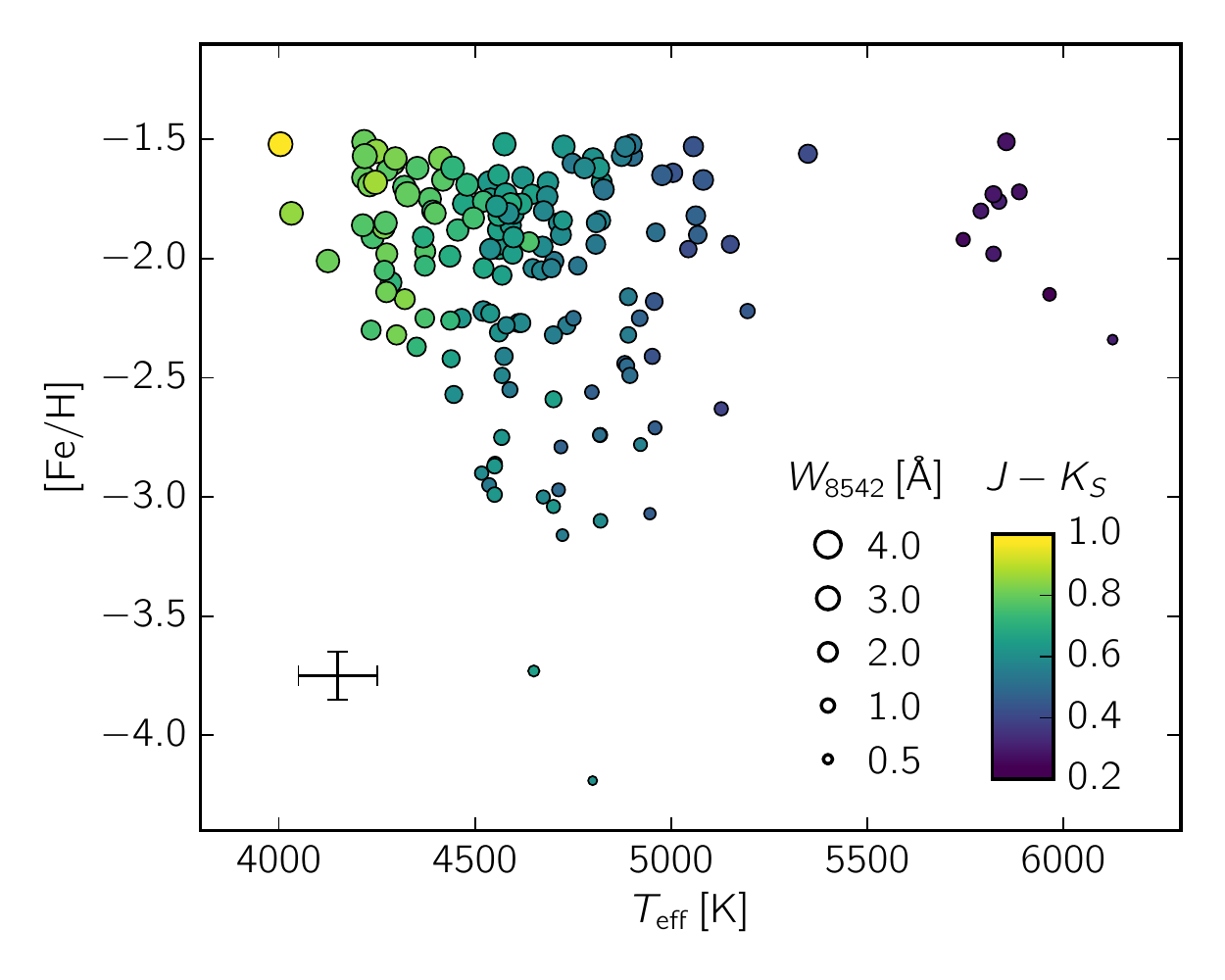}}
\caption{Sample of stars used as a training set for the Gaussian process model. Larger circles correspond to larger EW measurements. Darker colors mark bluer stars, brighter colors are redder. The typical uncertainty of the temperature and metallicity measurements are given by the error bar on the left.}
\label{fig_calibration_sample}
\end{figure}

In order to train the model, we need a set of stars for which the observables as well as the labels and all accompanying uncertainties are known. Optimally, such a set should cover the whole range of values of the observables as well as label values that we expect to see in the set of stars whose observables we will evaluate with the model. The majority of the stars in our calibration set (150 out of $\ncalibration$) were sourced from the catalog compiled by \citet{2013MNRAS.429..126R}. The selection in this catalog was initially based on RAVE observations so that all of the stars in the catalog have accompanying RAVE spectra. The set includes a mixture of metal-poor stars $(\met < -0.7)$ mostly from the red giant branch, but also some stars from the horizontal branch. About a quarter of the stars in the sample are main-sequence and main-sequence turn-off dwarfs. Stars in the sample come from all parts of the Galaxy although they are more often from the thick disk or the halo. The subset we used for training was limited to all stars below $\met=-1.5$ (150 out of 319). Our goal is to compute the metallicities of all the candidates below $\met=-2.0,$ but we wish to ensure that the model has enough range at the upper edge of this domain as well. We also excluded five cases where the RAVE spectra were problematic \citep[as recognized by][]{2012ApJS..200...14M}. On the very metal-poor side, the metallicities of the stars in this sample reach about $\met = -3.0$. To extend this limit even further, we added 9 stars from four other catalogs \citep{2013MNRAS.434.1681C, 2013A&A...551A..57H, 2014AJ....147..136R, 2013ApJ...771...67I} with the requirement that the label values were derived from high-resolution spectroscopy and that the stars were observed by RAVE. This brought the lower limit to $\met=-4.2$ by incorporating the star CD-38 245 discovered by \citet{1984ApJ...285..622B}. In all cases, we used values derived under LTE assumptions. The calibration sample is shown in Fig. \ref{fig_calibration_sample}. 

\subsection{Training the model}

Training the model requires optimizing the likelihood function in Eq. \ref{eq_gpcannon_lnlike}. Individual label covariance matrices in Eq. \ref{eq_label_cov} were not available in our calibration set source catalogs. We assumed the off-diagonal values of the label covariance matrix to be zero. This assumption is, of course, violated in reality, but it does not influence the calibration significantly. To simplify the calculation, we also assumed that all calibration set stars have equal uncertainties. We adopted the values $\Sigma_{T_\mathrm{eff}} = 100\,\mathrm{K}$ and $\Sigma_{\met} = 0.1\,\mathrm{dex}$. These values are close to the values reported by the authors of the catalogs.

Training was made by sampling the posterior generated from the likelihood in Eq. \ref{eq_gpcannon_lnlike} multiplied by a non-restrictive prior. After the sampler reached a stationary state, we used the parameters corresponding to the highest likelihood value as our best-fit estimate. This was done for each of the five observables, resulting in 20 parameters (three kernel parameters and one intrinsic scatter parameter per observable).

\section{Calibrated sample properties}
\label{sect_prop}

\begin{figure}
\centering
\resizebox{\hsize}{!}{\includegraphics{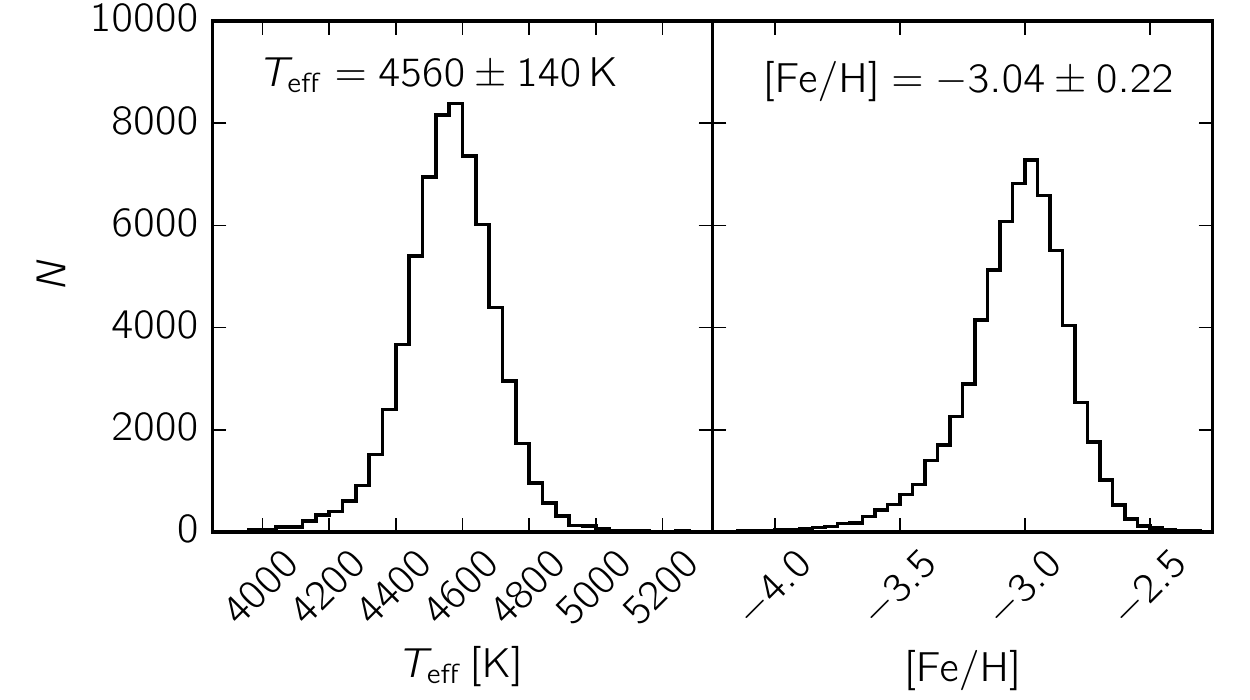}}
\caption{Posterior distributions of effective temperature and metallicity for the extremely metal-poor star whose spectrum is shown in Fig. \ref{fig_mp_spectrum}.}
\label{fig_label_posteriors}
\end{figure}

\begin{figure*}
\centering
\includegraphics[width=17cm]{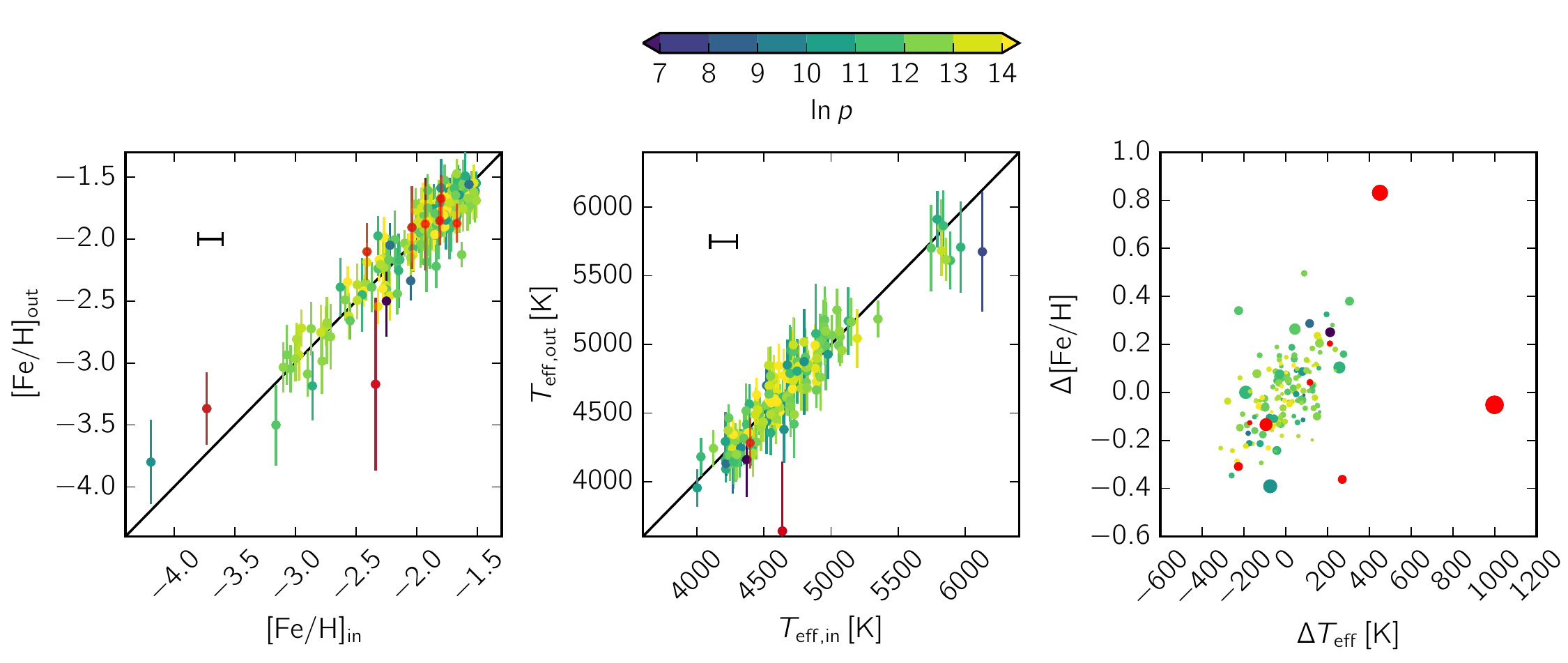}
\caption{Leave-one-out cross-validation of the training sample. The left and center panels show input vs. output values for $\met$ and $T_{\mathrm{eff}}$. The uncertainty of the input value is given by the error bar in each diagram. The right panel shows the differences between the input and the output values for both parameters, with larger circles representing solutions with higher total relative uncertainty (in both temperature and metallicity). Brighter colors correspond to the more reliable solutions with higher likelihood $\ln p$. Red symbols mark all solutions with multimodal posteriors for which a single value is a poor estimator and should therefore not be used. }
\label{fig_loo}
\end{figure*}

The best-fitting values of the kernel parameters and the intrinsic scatter were employed in the next stage of evaluating the observables (equivalent widths and colors) of candidate stars with unknown metallicities and temperatures. We used a sampler to generate a posterior distribution of the labels through likelihood in Eq. \ref{eq_label_likelihood} for each candidate star. At each iteration, the predictions were generated using Eqs. \ref{eq_label_mean} and \ref{eq_label_cov}. When computing the posterior distribution, we limited the values of temperature and metallicity by imposing a flat prior between $3000\,\mathrm{K}$ and $8000\,\mathrm{K}$ for the temperature and $-5.0\,\mathrm{dex}$ and $-1.3\,\mathrm{dex}$ for the metallicity. Posterior distributions of a typical extremely metal-poor star in the RAVE database are shown in Fig. \ref{fig_label_posteriors}. In most cases, the posterior distributions are smooth and normal-like.

\begin{figure*}
\centering
\includegraphics[width=17cm]{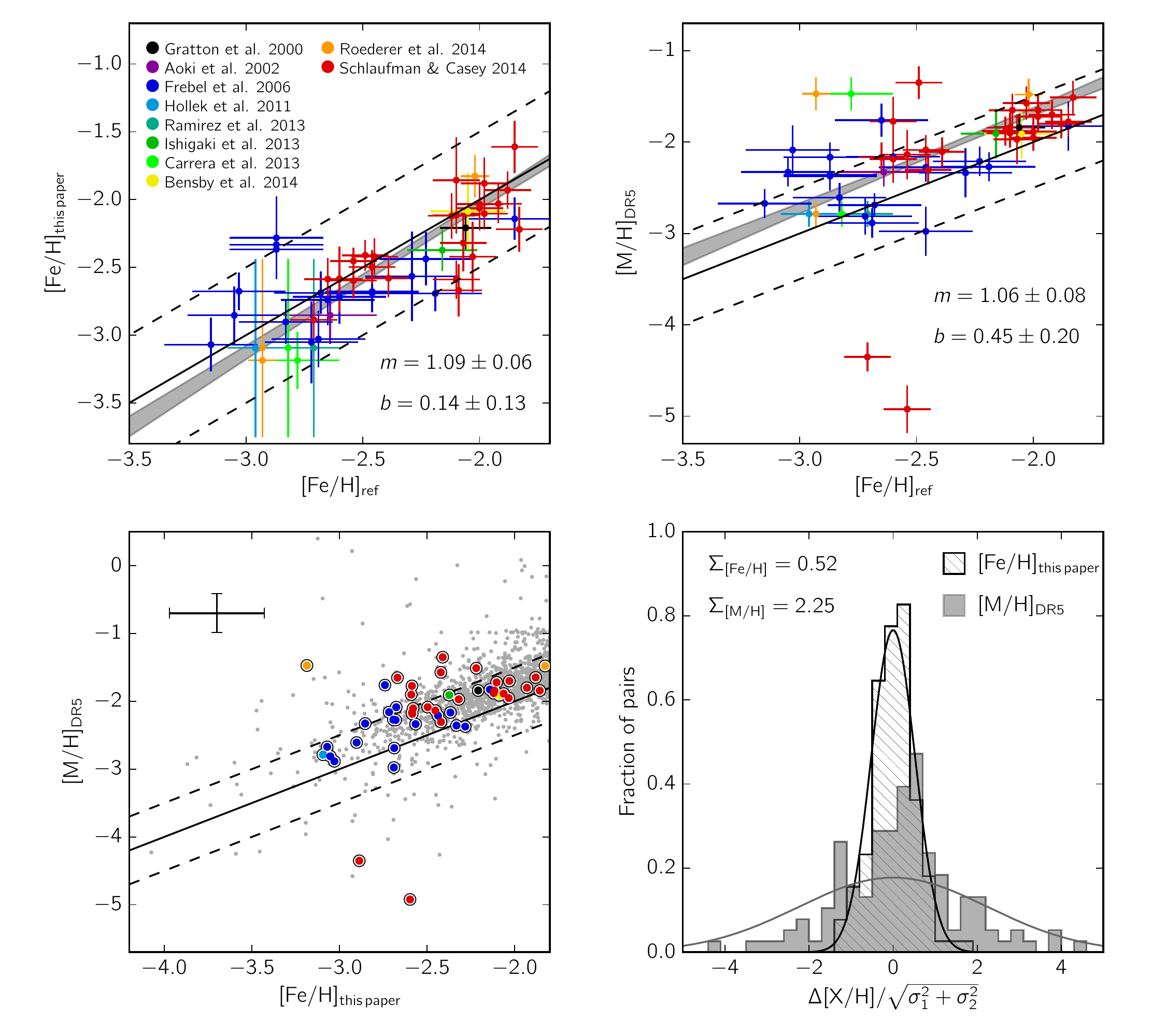}
\caption{Top left panel: comparison of the derived metallicity values to the literature values from various sources. The top right panel is similar, except that the vertical axis shows the values derived by the RAVE DR5 pipeline ($\mathrm{M}$ in the label denotes all elements heavier than $\mathrm{He}$). Reference data are from \citet{2000A&A...354..169G, 2002ApJ...580.1149A, 2006ApJ...652.1585F,2011ApJ...742...54H,2013ApJ...764...78R,2013ApJ...771...67I,2013MNRAS.434.1681C,2014A&A...562A..71B,2014AJ....147..136R, 2014ApJ...797...13S}. All reference metallicities were derived from spectroscopy except for those given by \citet[][dark blue]{2006ApJ...652.1585F}, which are photometric. The gray areas denote the $\pm 1\sigma$ linear fits to the data points. The line slope $(m)$ and intercept $(b)$ for each are given in the respective panels. The bottom left panel shows the relation between newly derived values and RAVE DR5 metallicities for candidates with $\met < -1.9\,\mathrm{dex}$ (individual error bars are not shown for clarity, but we show a representative error bar in the upper left corner). Note the difference in the metallicity range on both axes. The bottom left panel is similar to the top row in Fig. \ref{fig_gp_chi2}, except that it shows the pairwise differences in metallicity measurements between the repeated observations. The filled histogram represents DR5 values, while the hatched histogram shows the new values. The overplotted lines are Gaussians with widths measured from the data and indicated in the upper left corner of this panel.}
\label{fig_reffeh}
\end{figure*}

To verify the reliability of our model, a number of tests were performed. In the first, we performed a leave-one-out cross-validation to investigate how well the model predicts the temperature and metallicity values when the input from a star that is within the training domain is used as input. Iterating through the training set, we removed one star from the training set in each iteration and trained the model as described in the previous section. We then evaluated the values of the removed star on the freshly trained model and compared them to the known high-resolution determined values. This was repeated for every star in the training set. The results of this test are shown in Fig. \ref{fig_loo}. Both atmospheric parameters behave well across the whole range. Metallicity is recovered in the most metal-poor regime where only a few stars are used as a support in the training set. Even when the most metal-poor star in the training set is removed and the model is trained without it, its value still is evaluated within one standard deviation of the posterior distribution. As shown in the figure, input and output values of the temperature also correlate well. Since only a single star in the training set extends past the $6000\,\mathrm{K}$ mark, it is not to be expected that the model will be able to extrapolate beyond this point. This can be detected in the recovery of the turn-off group stars. However, the uncertainties also grow for temperatures that are close to or beyond this limit, which indicates that the model correctly responds to the observables that lie beyond the training set. In very few cases, the difference between the input and the output values becomes large (see the red points with largest uncertainties in the two diagrams in Fig. \ref{fig_loo}), but even in these cases, the model responds correctly by predicting a large uncertainty of the parameters. Generally, this test indicates that our uncertainties are slightly overestimated. As an additional test of the reliability of the determined parameters, we computed a statistic that measures the possible multimodality of the parameters' posteriors \citep[Dip test,][]{hartigan1985}. Cases that failed the test and where individual posterior distributions show a possible bimodal structure are marked with red dots in Fig. \ref{fig_loo}.

The second test of the method was performed by comparing the values of the derived parameters to the external catalogs for a handful of stars that were already known to be very metal-poor (these stars were not included in the training set). The results of the test are shown in Fig. \ref{fig_reffeh}. Metallicities derived using our pipeline correlate better with the reference values than the values derived using the RAVE DR5 pipeline, although the DR5 solution for many of these stars did not converge properly so the values are known to be unreliable. It should be noted that for a few stars whose metallicities were measured by multiple investigators, the values differ by as much as $0.3\,\mathrm{dex}$. The uncertainties derived by the DR5 pipeline also seem to be underestimated when compared to the newly derived counterparts. The bottom left panel in Fig. \ref{fig_reffeh} shows the comparison between our metallicity values and those derived by the DR5 pipeline. There is an apparent correlation, but also a significant scatter around them. The DR5 values are also shifted slightly upward with respect to the new values, but this is likely the consequence of choosing the LTE-derived metallicities in our training set, which are usually slightly lower than non-LTE-derived values.

Judging from the two tests, it seems that our model behaves well across the defined range in both temperature and metallicity, and it agrees well with the externally derived metallicity values.

As the third test, we compared the change in the derived values between the repeated observations of same stars. Ideally, the values should remain within the estimated error bars. From the bottom right panel of Fig. \ref{fig_reffeh} it is evident that the pairwise differences remain well within the uncertainties for the newly derived values. The DR5 pipeline metallicites, however, sometimes differ very significantly between the observations of the same star.

\begin{figure}
\centering
\resizebox{\hsize}{!}{\includegraphics{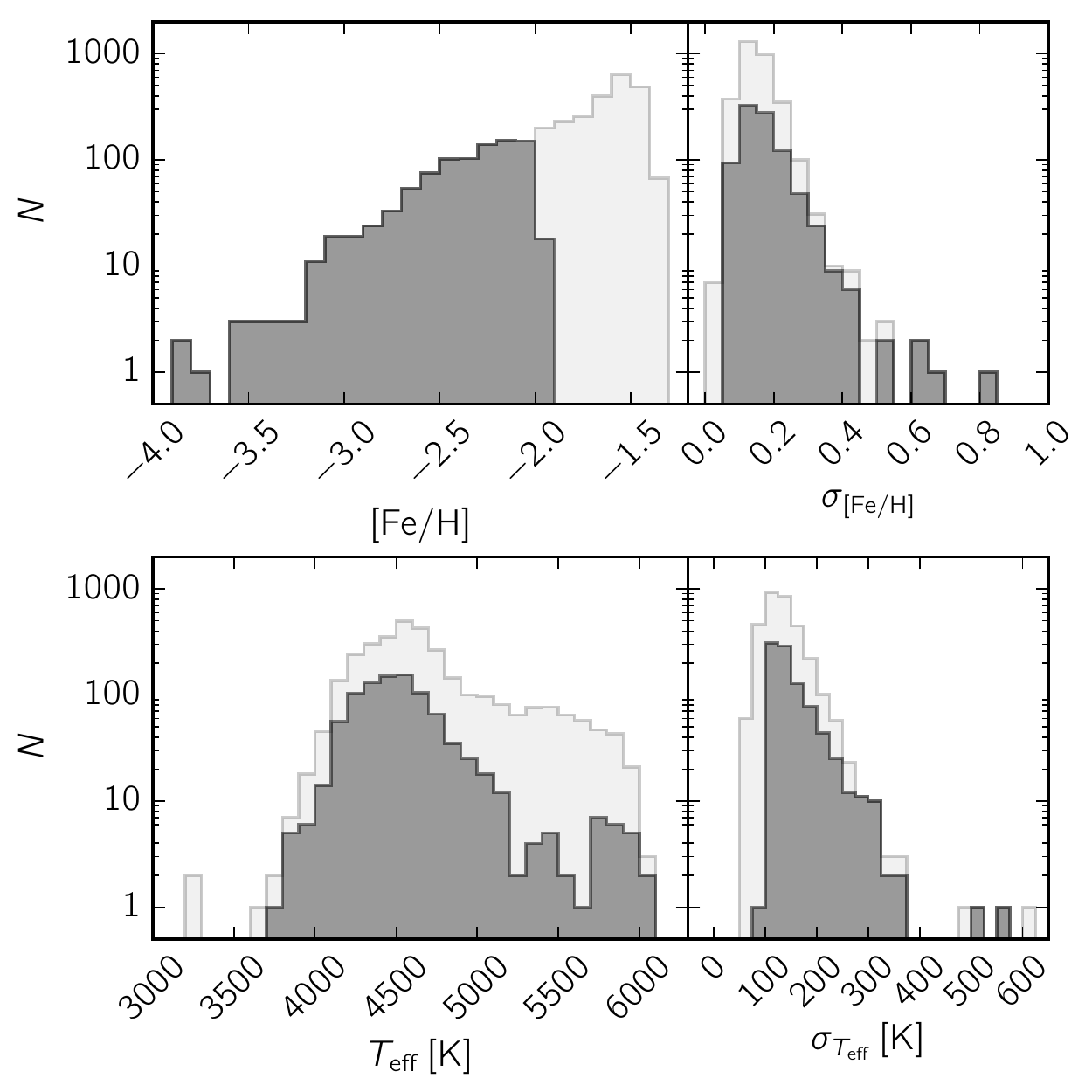}}
\caption{Metallicity distribution function and the distribution of the metallicity uncertainties (top), and the same for effective temperature (bottom). The means of the individual posteriors were used to construct the histogram on the left side. The right side histograms show the standard deviations of the posteriors. The light gray histograms include all candidate stars, and the dark gray histograms include all cases with a 50\% probability for the computed metallicity to be at least$\text{}$  below $\met=-2\,\mathrm{dex}$.}
\label{fig_label_dists}
\end{figure}

\section{Results}
\label{sect_results}

Out of $\nmpcandidate$ candidates, we present $\nvmpcandidate$ stars with a probability of at least $50\,\%$  to be very metal-poor (the median of the metallicity posterior distribution is lower than $-2\,\mathrm{dex}$) and $\nempcandidate$ with a probability
of at least $50\,\%$  to have $\met < -3\,\mathrm{dex}$. Five hundred and four stars have a probability of at least $95\,\%$  to be very metal-poor. There are no new discoveries with metallicities below $\met = -4\,\mathrm{dex}$, which means CD -38 245 remains the most metal-poor star observed by RAVE. The metallicity (MDF) and temperature distribution along with the uncertainty distributions are shown in Fig. \ref{fig_label_dists}. The MDF has a strong peak around $\met = -1.5\,\mathrm{dex}$, which is erroneous since many of the candidates are more metal-rich than our upper limit of the training set, which consequently pushes all these candidates toward this limit. To avoid these, we only accepted cases with a median metallicity below $-2\,\mathrm{dex}$ (represented by the darker shaded histograms in Fig. \ref{fig_label_dists}). The typical uncertainty in the metallicity is around $\sigma_{\met}=0.2\,\mathrm{dex}$. From the temperature distribution it is evident that the large majority of the very metal-poor candidates come from the giant branch and only a handful are turn-off stars. The reason for this is that giants are farther away than equally bright turn-off stars, which places them in the more metal-poor Galactic environment, therefore this is a product of the selection bias \citep{2016arXiv161100733W}. As expected, the ratio of rejected candidates (with $\met > -2\,\mathrm{dex}$) is much higher among the turn-off stars ($T_\mathrm{eff} \gtrsim 5500\,\mathrm{K}$). The typical uncertainty of the temperature is around $140\,\mathrm{K,}$ and the majority of the stars $(94\,\%)$ have temperature uncertainties below $200\,\mathrm{K}$.

\section{Conclusions}
\label{sect_conc}

A complete search to identify the most metal-poor stars observed by the RAVE survey was carried out. This study is a continuation of the work done by \citet{2010ApJ...724L.104F}, now using the full DR5 sample. We employed a three-stage method to identify the metal-poor candidates, to evaluate their $\caii$ triplet line equivalent widths, and to convert them into $\met$ measurements by also incorporating the color information provided by the 2MASS and WISE surveys.

The main difference in the creation of the candidate sample compared to the previous RAVE metal-poor study is that only spectral information was used, and we did not rely on any externally derived parameter, such as effective temperature. The selection was performed by projecting the spectra on a low-dimensional manifold using the t-SNE algorithm and then making a selection in a particular part of the manifold that is mostly populated by very metal-poor stars. In this way, we created a list of $\nmpcandidate$ metal-poor candidates. We estimate that this selection is very close to complete (i.e., almost no very metal-poor stars are left out). This prediction is based on the manual examination of the whole t-SNE projection and the assessment that no other regions of it are populated by metal-poor star spectra.

In the second stage we computed the equivalent widths of the $\caii$ triplet lines using a flexible model incorporating Gaussian processes to account for the correlations in the residuals. While fast and mature codes for computing equivalent widths of lines in stellar spectra such as ARES \citep{2015A&A...577A..67S} work well on spectra where correlated noise can be neglected, its presence can severely skew the equivalent width measurements if not accounted for. The comparison between the equivalent width values computed using a standard $\chi^2$ approach and noise-modeled values shows significant differences between them. The subset of stars for which we have multiple spectra per star from repeated observations revealed that the $\chi^2$ approach yields skewed face values and underestimated uncertainties. On the other hand, the Gaussian processes model gave us more reliable estimates and realistic uncertainties.

As the last stage, we constructed a method to convert the equivalent width measurements and the color information provided by the 2MASS and WISE surveys into metallicity values. In addition to the candidate sample, we have $\ncalibration$ stars for which abundances derived from high-resolution observations are available. Instead of fitting the equivalent width--color--metallicity relation on this set using a predefined function, we opted for an alternative approach. We designed a model similar to \textit{The Cannon} to train the model on this subset, again employing Gaussian processes to give the model the required flexibility. After the model was trained, we exploited the predictive power of Gaussian processes to evaluate the equivalent width and color information and to produce metallicity estimates and their uncertainties for candidate stars. A leave-one-out cross-validation test and comparison to literature values confirmed that our estimates have very little bias and the uncertainties are trustworthy. This way, we discovered $\nvmpcandidate$ new likely very metal-poor stars, extending the previous RAVE metal-poor star study by \citet{2010ApJ...724L.104F}. Thirty-nine candidates are likely to be extremely metal-poor. We intend to further explore this sample by acquiring follow-up observations using higher-resolution instruments. In addition to this, because these stars are bright, the distances that will be provided by the \textit{Gaia} satellite for many of the identified very metal-poor stars will enable the computation of their orbits. The described approach is also easily adapted to be used on other $\caii$ data sets (e.g., future Gaia data releases), or trained on a different spectral region (SDSS, LAMOST, etc.) and/or color information.

\begin{acknowledgements}
We would like to thank Borja Anguiano and Else Starkenburg for carefully reading the paper and providing us with comments that helped to improve the paper. We also thank the anonymous referee for useful comments. Funding for RAVE has been provided by the Australian Astronomical Observatory; the Leibniz-Institut fuer Astrophysik Potsdam (AIP); the Australian National University; the Australian Research Council; the French National Research Agency; the German Research Foundation (SPP 1177 and SFB 881); the European Research Council (ERC-StG 240271 Galactica); the Istituto Nazionale di Astrofisica at Padova; The Johns Hopkins University; the National Science Foundation of the USA (AST-0908326); the W. M. Keck Foundation; the Macquarie University; the Netherlands Research School for Astronomy; the Natural Sciences and Engineering Research Council of Canada; the Slovenian Research Agency; the Swiss National Science Foundation; the Science \& Technology Facilities Council of the UK; Opticon; Strasbourg Observatory; and the Universities of Groningen, Heidelberg and Sydney.
The RAVE web site is at https://www.rave-survey.org. This research made use of CDS tools SIMBAD and VizieR, NASA ADS and python packages \texttt{numpy}, \texttt{spicy}, \texttt{matplotlib}, \texttt{pandas}, and \texttt{pyspeckit}.
\end{acknowledgements}

\bibliographystyle{aa}
\bibliography{rave_very_metal_poor.bbl}

\end{document}